\documentclass[prd,aps,floats,floatfix,superscriptaddress,preprintnumbers,
showpacs,eqsecnum,nofootinbib,twocolumn]{revtex4}
\usepackage{amsfonts,
amsmath,epsfig,amssymb,
latexsym,color,graphicx,array,
theorem,mathrsfs,subfigure,bm,float}

\definecolor{red}{rgb}{1,0,0}
\definecolor{blue}{rgb}{0,0,1}
\definecolor{green}{rgb}{0,1,0}

\newcommand{\ma}[1]{\mbox{$\mathcal{#1}$}}

\newcommand{\D}{{\rm d}}
\newcommand{\ti}{\tilde}

\newcommand{\vect}[1]{\!\!\!\mbox{ \boldmath $#1$}}
\newcommand{\calhR}[1]{\raisebox{2ex}{\tiny ({\em h})}\hspace{-0.8em}{\ma R}}

\begin{document}

\title{
Black Hole in the Expanding Universe
with Arbitrary Power-Law Expansion
}


\author{Kei-ichi {\sc Maeda}}
\email{maeda@waseda.jp}
\address{Department of Physics, Waseda University, 
Okubo 3-4-1, Shinjuku, Tokyo 169-8555, Japan}
\address{Waseda Research Institute for Science and Engineering,
Okubo 3-4-1, Shinjuku, Tokyo 169-8555, Japan}
\author{Masato {\sc Nozawa}}
\email{nozawa@gravity.phys.waseda.ac.jp}
\address{Department of Physics, Waseda University, 
Okubo 3-4-1, Shinjuku, Tokyo 169-8555, Japan}


\date{\today}

\begin{abstract} 
We present a time-dependent and spatially inhomogeneous solution that interpolates the extremal Reissner-Nordstr\"om (RN) black hole and the 
Friedmann-Lema\^itre-Robertson-Walker (FLRW) universe with arbitrary power-law expansion. It is an exact solution of the $D$-dimensional
 Einstein-``Maxwell''-dilaton system, where two Abelian gauge fields couple to the dilaton with different coupling constants, and the 
dilaton field has a Liouville-type exponential potential.  It is shown that the system satisfies the weak energy condition. 
The solution involves two harmonic functions on a $(D-1)$-dimensional Ricci-flat base space. In the case where the harmonics have a single-point source on the Euclidean space, we find that the spacetime describes a spherically symmetric charged black hole in the FLRW universe, which is characterized by three parameters: the steepness parameter of the dilaton potential $n_T$, the U$(1)$ charge $Q$, and the ``nonextremality'' $\tau $. In contrast with the extremal RN solution, 
the spacetime admits a nondegenerate Killing horizon unless these parameters are finely tuned. The global spacetime structures are discussed in detail. 
\end{abstract}


\maketitle

\section{Introduction}

Black holes could have formed in the early stage of the universe as a
consequence of primordial density fluctuation~\cite{PBH1,PBH2}. 
Since these primordial black holes (PBHs) 
are of the order of the horizon scale mass,  
they span a considerably broad mass spectrum ranging  
from the Planck mass ($10^{-5}$g) to $10^5 M_{\odot}$, or even much
larger. Hence, their presence could leave diverse physical imprints 
throughout the cosmic history (see~e.g.,~\cite{PBHs} and references therein).  
The PBHs with $M>10^{15}$g survive until the present epoch, so that    
they are plausible candidates of cold dark matter, and likely 
origin of supermassive black holes and/or the sources
of gravitational waves. 
The PBHs with mass $M\sim 10^{15}$g are now evaporating via 
Hawking evaporation~\cite{Hawking1974}. 
Such PBHs emit quanta of order 100MeV, which  
could also contribute to the cosmological $\gamma $-ray background and 
generate $\gamma $-ray bursts.   
While, the PBHs with mass smaller than 
$10^{15}$g that have completely evaporated during a first second of big bang
could generate large amount of entropy of the universe. 
Thus, the number density of small PBHs could place 
strong constraints on the big-bang nucleosynthesis. 
In these contexts, studying the formation and the evaporation process of 
black holes 
in the expanding universe are of great significance as a probe of 
the early
universe, high energy physics, and quantum phase of gravity.

Apart from the above astrophysical interests, 
a study of black holes in the expanding universe has been 
a principal stirring subject in general relativity. 
Black holes have played a central role in general relativity, 
since they encode the essential characteristics of the gravity theory,
viz,  the nonlinearity and highly curved spacetime.
Various kinds of physical and geometrical properties of black holes have been 
clarified thus far.   The most enduring achievements among these is the  
uniqueness theorem of black holes~\cite{Carter,uniqueness}, according to which 
isolated black holes in equilibrium states--these states occur if sufficiently long time
passed after the  gravitational collapse of massive stars--necessarily belong to 
the Kerr family. This means that stationary black holes are 
completely characterized by two conserved charges (mass and angular
momentum)  without any additional ``hairs.'' 
By virtue of this theorem, stellar-sized black holes with
futile ambient sources are well approximated by Kerr black holes. To the contrary, 
in the non-isolated and dynamical background,   
a variety of black holes equipped with much richer properties 
are expected. However,   
when one attempts to obtain the black hole spacetime 
in the dynamical background, a serious difficulty arises. 
If a black hole is put on the homogeneous and isotropic 
FLRW universe  on which the standard
cosmological  scenario lays the foundation, the background 
universe will become inhomogeneous, and at the same time 
the black hole will continue to grow and/or deform by swallowing ambient
matters.
Hence, the fact that the spacetime is time-evolving and spatially
inhomogeneous enforces us to 
solve nonlinear partial differential equations for the geometry, as well
as for the matter fields.

The exact  black-hole solutions in the FLRW universe found 
hitherto have enjoyed high degrees of symmetry. Among them
are the Einstein-Straus model~\cite{Einstein:1945id} 
(a black hole in the ``Swiss-Cheese Universe''),  the 
Schwarzschild-de Sitter (SdS) and the Reissner-Nordstr\"om-de Sitter (RNdS)
black hole.  
Since these solutions maintain the equilibrium due to the timelike Killing 
field,  
they are unlikely to capture the envisaged dynamical
fact in realistic situations.  
In recent years,  Sultana and Dyer have exploited the conformal technique and
have constructed a dynamical black hole~\cite{SultanaDyer} which is
asymptotically Einstein-de Sitter universe and   
conformally related to the Schwarzschild solution.    
Unfortunately, the Sultana-Dyer solution suffers from violating energy 
conditions. 
Meanwhile, if one imposes a self-similarity (characterized by a 
homothetic Killing field), it has been proved that a spherically symmetric 
black hole cannot exist in the asymptotically decelerating  
FLRW universe~\cite{HMC}. These examples illustrate the difficulty in 
constructing  a regular black hole immersed in the FLRW universe 
with desirable physical properties.

Recently,  a time-dependent ``black hole 
candidate'' was obtained
via the dimensional reduction of  the dynamically intersecting branes in
11-dimensional (11D) supergravity~\cite{MOU}. 
The solution appears to behave like a charged black hole for small
radii, while for large radii it approaches to the FLRW universe filled
with stiff matter fluid.
In the previous paper~\cite{MN}, we elucidated the detailed spacetime
structure and established that the solution indeed describes a charged 
black hole
that approaches to the flat FLRW  universe. 
It is shown that the metric is an exact solution in
Einstein-``Maxwell''-dilaton system with four kinds of U$(1)$-fields (three
of them are degenerate) coupled to the dilaton, thereby the system  
satisfies the dominant energy condition.  
Although the BPS black hole plays the leading part in string
theory, what is intriguing us 
is that the solution is {\it not} extremal, i.e., the Hawking
temperature does not vanish.  
Despite the little astrophysical concern because of the brane charges, 
the intersecting brane picture will enable us to obtain 
a variety of dynamical black holes with physical matter sources. 
In~\cite{GMII},  Gibbons and one of the present authors
found the ``black hole candidate'' that asymptotically 
looks like an FLRW universe with arbitrary power-law expansion 
by introducing an exponential potential. 
The solution found in~\cite{GMII} 
reduces to the solution in~\cite{MN,MOU} 
as a special case and is expected to describe a black hole  
in the expanding universe. 
As lessons from the McVittie solution~\cite{Mcvittie1933,Nolan,Carrera:2009ve}, 
a great deal of care is needed in order to obtain the global spacetime
structure of a time-evolving and spatially inhomogeneous
solution.\footnote{In contrast to the claims in~\cite{Nolan}, 
it is recently argued that McVittie's solution includes
a regular black-hole horizon if the positive cosmological constant dominates the
universe at late time~\cite{Kaloper:2010ec}. 
} 
In this paper, following the previous study~\cite{MN} we intend to
clarify the spacetime structure of their solution, which has been an
open issue in~\cite{GMII}.

There are several motivations for studying black holes in 
the Einstein-``Maxwell''-dilaton system with an exponential potential.
In the light of string theory,  
the dilaton and the form fields are elementary constituents in the
theory, and the exponential potential of a dilaton naturally arises in various 
contexts: the excess central charge (the ``noncritical'' string theory)~\cite{noncritical_string_theory}, 
the massive IIA supergravity~\cite{massiveIIA}, 
the Kaluza-Klein compactification of a 
curved internal space~\cite{Kaluza-Klein_compactification}, 
the vacuum expectation value of the four-form field
strength~\cite{VEV_form}, and so on. 
From the general relativistic point of view,  
these fields obey suitable energy conditions. Thus, it is 
interesting to see if the black hole exhibits a peculiar aspect
under a realistic circumstance.   
In addition, since the exponential potential may drive the 
power-law inflation~\cite{Lucchin:1984yf}, this could give a  
profound implication for the background geometry of PBHs during inflation. 
Furthermore, by generalizing the present solution to include multiple
black holes one is able to discuss the collision and 
coalescence of black holes,    
providing a valuable arena to test the cosmic censorship 
conjecture~\cite{Penrose:1969pc}.

This paper constitutes as follows. 
In the ensuing section, we present a $D$-dimensional solution as a
simple and honest generalization of the solution given in~\cite{GMII}.  
We discuss the matter fields and singularities of the solution in
Section~\ref{sec:matter}, where the system is shown to satisfy the
suitable energy conditions. Section~\ref{sec:spacetime} develops our main
argument on the spacetime structure. 
Analysis of  the near-horizon geometry and null geodesic motions
leads to the conclusion that the present spacetime can admit a 
regular event horizon with constant circumference radius, if the 
parameters of the solution are chosen appropriately.  
Conclusions and future outlooks are summarized in Section~\ref{sec:summary}.

We shall work in units $c=\hbar =1$ with $\kappa^2 =8\pi G$ and follow
the standard curvature conventions 
${\ma  R^\mu}_{\nu\rho\sigma}V^\nu:=2\nabla_{[\rho }\nabla_{\sigma ]}V^\mu$,
$\ma  R_{\mu \nu }:={\mathcal R^{\rho }}_{\mu \rho \nu }$
and 
$\ma  R:=\ma  R^\mu_\mu $. Since the circumference radius will be
denoted by $R$, the script notation is used for curvature tensors throughout
the paper.

\section{Preliminaries}
\label{sec:preliminaries}

The solution describing a charged black hole in
the FLRW universe is derived in~\cite{MOU} from the dimensional reduction of
dynamically intersecting M2/M2/M5/M5-branes in 11D supergravity.
Viewed from 4D inhabitants, 
the solution satisfies the field equations in Einstein-``Maxwell''-dilaton
system  with particular couplings. The gauge fields and the scalar field
arise via toroidal compactification of extended directions of branes
in 11D supergravity.
The massless scalar field is responsible for the 
stiff-fluid dominant universe.   
This solution was generalized in~\cite{GMII} in such a way that the 
background FLRW cosmology obeys the arbitrary power-law expansion
by introducing the Liouville-type potential.

In this paper, a more general class of solutions are to be considered. 
We shall begin by the $D$-dimensional Einstein-``Maxwell''-dilaton system, 
in which two types of U$(1)$-fields couple to the dilaton with different
couplings, and the dilaton has a Liouville-type exponential potential. 
To be specific, the action is described by
\begin{align}
S=& \int \D^D x\,\sqrt{-g}\,
\left[\frac{1}{2\kappa^2 } \ma R-\frac 12 \left(\nabla_\mu  \Phi \right)
\left(\nabla^\mu \Phi\right)
\right. \nonumber \\
& \left.-V(\Phi)-\frac 1{16\pi} \sum_{A=S, T}n_A 
e^{\lambda _A\kappa \Phi}
F_{\mu \nu }^{(A)} F^{(A)\mu \nu }\right]\,,
\label{action}
\end{align}
with
\begin{align}
V(\Phi) =V_0 \exp(-\alpha \kappa \Phi )\,.
\label{V}
\end{align}
Here, $\alpha ~(\ge 0)$ is a dimensionless constant corresponding 
to the steepness of the potential. 
We have introduced degeneracy factors, $n_A~(\ge 0)$, 
of two U$(1)$ fields for later convenience (which may be
absorbed into the definition of $F^{(A)}_{\mu \nu }$).
This type of action with
a single U$(1)$ field has been discussed in 
some situation, e.g. 
a black hole collision in 
the $D=4$ theory with $\alpha =-\lambda_T=2$~\cite{HH}
(see also \cite{MS}).

The above action~(\ref{action})
yields the field equations
for the metric, the dilaton, and two U$(1)$ fields as,
\begin{align}
& \ma R_{\mu \nu }- \frac 12 \ma R g_{\mu \nu }=\kappa^2 
\left(T^{(\Phi)}_{\mu \nu }+T^{(\rm em)}_{\mu \nu }\right)\,, 
\label{Einsteineq}\\
&\nabla^\mu \nabla _\mu \Phi -\frac{\D V}{\D \Phi}-\frac{1}{16\pi}
\sum_An_A\lambda_A \kappa e^{\lambda_A \kappa \Phi} F_{\mu \nu }^{(A)}
 F^{(A)\mu \nu }=0\,,\label{dilatoneq} \\
&\nabla_\nu \left(e^{\lambda_A \kappa \Phi }F^{(A)\mu \nu }\right)=0\,,
\label{Maxwelleq}
\end{align} 
where 
\begin{align}
 T_{\mu \nu }^{(\Phi)} &=\left(\nabla_\mu \Phi \right)\left(\nabla_\nu
 \Phi \right)-\frac 12g_{\mu \nu }\left[
\left(\nabla^\rho \Phi \right)
\left(\nabla_\rho \Phi \right)+2V\right] \,,
\label{SEtensor_Phi} \\
T_{\mu \nu }^{\rm (em)} &= 
\sum_A \frac{n_A e^{\lambda_A \kappa \Phi }}{4\pi }
\left(F_{~\mu \rho }^{(A)}{F^{(A)\rho }_{~\nu} }-\frac 1 4g_{\mu \nu }
F_{\rho \sigma }^{(A)}
 F^{(A)\rho \sigma  }
 \right)\,.
\label{SEtensor_em}
\end{align}
Inspired by the solutions given in~\cite{MOU,GMII}, 
let us specify the parameters $(n_A, \lambda_A, \alpha )$
as 
\begin{align}
& n_T+n_S=\frac{2(D-2)}{D-3}\,, 
\\
&\lambda_T=\alpha =2\sqrt{\frac{(D-3)n_S}{(D-2)n_T}}\,,
\\
&\lambda _S=-2\sqrt{\frac{(D-3)n_T}{(D-2)n_S}}\,,
\end{align}
for which 
\begin{align}
0\le n_T\le 4 \,,~~ 
0\le n_S \le 4 \,,~~
\lambda_T\ge 0 \,,~~
\lambda_S\le 0 \,.
\end{align}
The constants $n_T$ and $n_S$ may take natural number only for $D=4, 5$,
 in which
case they are related to the number of time-dependent and static branes
in 11D supergravity~\cite{MN,MOU}.
In the present case, however, $n_T$ and $n_S$ are not
necessarily integers, hence they do not have such a meaning.

With the above choice of parameters, 
we shall look for a spatially-inhomogeneous and time-dependent  solution
of the following form, 
\begin{align}
\D s^2=-\Xi^{D-3}\D t^2+\Xi ^{-1} \,
\D \sigma_{D-1}^2 \,,
\label{metric}
\end{align}
with
\begin{align}
&\Xi := \left[\left(\frac{t}{t_0}+ \bar H_T \right)^{n_T}
\left(1+ \bar H_S\right)^{n_S}\right]^{-1/(D-2)}\,,
\label{Xi}
\\
&\D \sigma_{D-1}^2 :=h_{IJ}(x) \D x^I \D x^J 
\,,
\end{align}
where $t_0$ is a constant with dimension of time, and 
$ \bar H_T(x)$ and $\bar H_S(x)$ are 
functions of $x^I$ on the  $(D-1)$-dimensional base space 
$\D \sigma_{D-1}^2$.

Assuming that 
the base space is Ricci-flat,
\begin{align}
^h{\ma R}_{IJ}=0,
\end{align} 
where $^h{\ma R}_{IJ}$ is the Ricci curvature of the spatial metric $h_{IJ}$, 
and $\bar H_T(x)$ and $\bar H_S(x)$ are  harmonic functions
of the base space  $\D \sigma_{D-1}^2$, 
\begin{align}
^h\Delta 
\bar H_T=0\,,\quad 
^h \Delta 
\bar H_S=0
\,,
\end{align} 
and if $t_0$ is tuned to be
\begin{align}
 t_0^2=\frac{n_T (n_T-1)}{4\kappa^2 V_0}\,, 
\end{align}
we find that the metric~(\ref{metric})
solves the field equations 
(\ref{Einsteineq})--(\ref{Maxwelleq})
provided that the dilaton and the electromagnetic potentials
$A_\mu ^{(A)}$ defined by
\begin{align}
F_{\mu \nu }^{(A)}=\nabla_\mu A_\nu ^{(A)}-\nabla_\nu A_\mu ^{(A)}
\,,
\end{align} 
are given by
\begin{align}
 \kappa \Phi &=\frac{1}{2}\sqrt{\frac{(D-3)n_Tn_S}{D-2}}
\, \ln
 \left(\frac{H_T}{H_S}\right)\,,
\label{Phi}\\
\kappa A^{(T)}_t&=\sqrt{2\pi} H_T^{-1}+a^{(T)}(t)\,,
\\
\kappa A^{(S)}_t&=\sqrt{2\pi} H_S^{-1}+a^{(S)}(t)\,,
\end{align}
where 
\begin{align}
H_T={t\over t_0}+\bar H_T
\,,\quad 
H_S=1+\bar H_S
\,,
\end{align}
and 
$a^{(T)}(t)$ and $a^{(S)}(t)$ are arbitrary functions of $t$.

Bearing the cosmological application in mind, 
we set $h_{IJ}$ to be a flat Euclidean metric, $h_{IJ}=\delta_{IJ}$, 
in what follows. 
Note that the action is not invariant under
electromagnetic duality transformation except the $n_T=1$ case.  
We shall thence focus on the electrically charged solution
throughout the article.

Let us consider the case where the harmonics  represent
a system of $N$ point-sources,
\begin{align}
\bar H_T&=\sum_{i=1}^N {Q^{(i)}_T\over |\,\vect{x}-\vect{x}_{(i)}|}
\,,
\label{HT}\\
\bar H_S&=\sum_{i=1}^N {Q^{(i)}_S\over |\,\vect{x}-\vect{x}_{(i)}|}
\,,
\label{HS}
\end{align}
where $Q^{(i)}_T$ and $Q^{(i)}_S$ are charges at the points 
$\vect{x}_{(i)}$~($i=1,\cdots,N$), respectively.
It follows that the case of $n_T=0$ reduces to the higher-dimensional
Majumbdar-Papapetrou solution~\cite{Hartle:1972ya}, 
while the case of $n_S=0$ describes the 
higher-dimensional Kastor-Traschen solution~\cite{KT,London}.
Henceforth, we do not  consider these special cases unless otherwise stated,
 i.e., 
we assume $n_T\ne 0$ and $n_S\ne 0$.  
For $D=4, n_T=1$, one recovers the solution in~\cite{MOU}, which has
been shown to describe a charged black hole in the FLRW universe
when each harmonic function has a single point source at 
the origin~\cite{MN}.

Supposed $(t/t_0)>0$,  let us transform to the new time 
coordinate $\bar t$ defined by 
\begin{align}
 \frac{\bar t}{\bar t_0}=\left(\frac{t}{t_0}\right)^
{{(D-3)\over 2(D-2)}n_S} ~~ {\rm with}~~~ 
\bar t_0=\frac{2(D-2)}{(D-3)n_S}\,t_0\,, 
\label{tbar}
\end{align}
in terms of which 
we can cast the metric~(\ref{metric}) into the form
\begin{align}
\D s^2 =-\bar \Xi^{D-3} \D \bar t^2+{a^2 }{\bar \Xi }^{-1}
\delta_{IJ} \D x^I \D x^J \,,
\label{metric_FRW}
\end{align}
where 
\begin{align}
\bar \Xi =\left[
\left(1+\frac{\bar H_T}{a^{2(D-2)/n_T}}\right)^{n_T}
\left(1+\bar H_S\right)^{n_S}
\right]^{-1/(D-2)}\,,
\end{align}
and 
\begin{align}
a=\left(\bar t\over \bar t_0\right)^p\,,  ~~ {\rm with}~~~ 
p=\frac{n_T}{(D-3)n_S}\,.
\label{scale_factor}
\end{align}
Since we imposed the boundary condition such that 
the harmonics $\bar H_T$ and $ \bar H_S$
fall off as $r:=\sqrt{\sum_I (x^I)^2}\to \infty $ [Eq.~(\ref{HT}) and (\ref{HS})],
the metric~(\ref{metric_FRW}) approaches in the limit $r\to \infty $ to the 
$D$-dimensional flat FLRW spacetime,
\begin{align}
\D s^2_{r\to \infty } =-\D \bar t^2 +a^2 \delta_{IJ}\D x^I \D x^J\,.
\end{align}
The new coordinate $\bar t$ is found to measure the proper time at infinity. 
Looking at the behavior of the scale factor $a\propto \bar t^p$, 
one can recognize that 
the asymptotic region of the spacetime is
the FLRW universe  filled by a fluid  with the  equation of state
\begin{align}
 P=w \rho\,,  ~~ {\rm with}~~~ 
w={2(D-3)n_S\over (D-1)n_T}-1
\,.\label{EOS}
\end{align}
It turns out that the 
parameter $n_T$ (or $n_S$) is associated to the expansion 
law of the universe (\ref{scale_factor}). 
Notably, we can obtain an accelerating universe $(p\ge 1)$ 
by setting $n_T\ge 2$ or equivalently $n_S\le 2/(D-3)$. 
In particular, the exponential expansion (the de Sitter universe)
is understood to be $p\to \infty ~(n_S\to 0)$.  
Figure~\ref{fig:FRW} depicts the conformal diagrams of the  
FLRW universe. The asymptotic regions of the 
present spacetime~(\ref{metric}) resemble the corresponding 
shaded regions in Figure~\ref{fig:FRW}.

\begin{widetext}
\begin{center}
\begin{figure}[h]
\includegraphics[width=14cm]{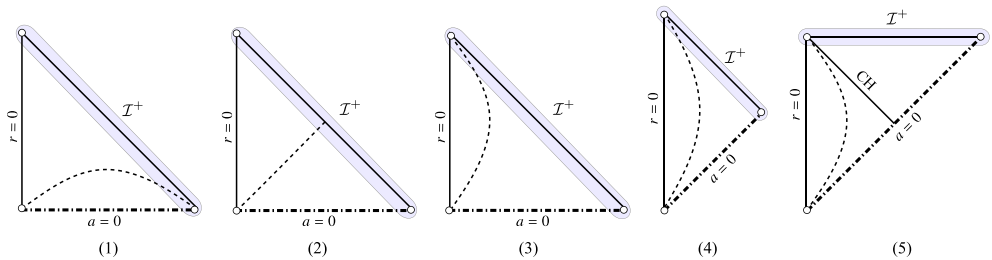}
\caption{Conformal diagrams of a flat FLRW universe 
$a = (\bar t/\bar t_0)^p$ for (1) $0<p<1/2$, 
(2) $p=1/2$, (3) $1/2<p<1$, (4) $p=1$ and (5) $p>1$. 
The dotted and dotted-dashed lines denote the trapping horizon,
$r_{\rm TH}(\bar t)=(\D a/\D \bar t)^{-1}$, and the big-bang
 singularity at $a=0$, respectively. 
The cases (2) and (4) correspond
 respectively  to the radiation-dominant universe $P=\rho/(D-1)$ and the    
marginally accelerating universe driven by the curvature term 
$\rho \propto a^{-2}$.  
The cosmological horizon,
$r_{\rm CH}(\bar t)=(p-1)^{-1}\bar t_0 (\bar t/\bar t_0)^{1-p}$, is 
abbreviated to CH, which exists only in the strictly accelerating 
case ($p>1$). 
The shaded regions corresponding to $r\to \infty $ approximate our original 
spacetime.
}
\label{fig:FRW}
\end{figure}
\end{center}
\vskip -.5cm
\end{widetext}

\begin{figure}[t]
\begin{center}
\includegraphics[width=9cm]{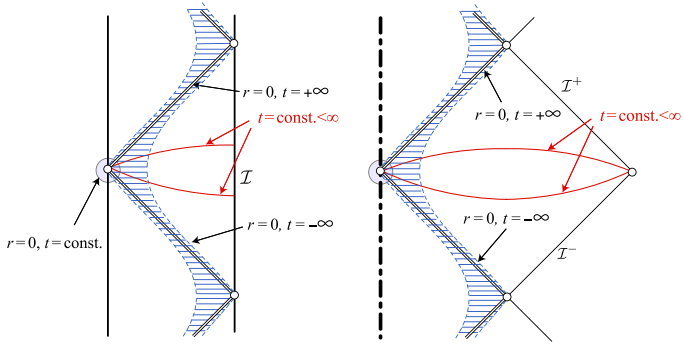}
\caption{Typical ``throat'' geometries for 
the AdS$_2\times {\rm S}^{D-2}$ spacetime (left) and the extremal RN spacetime 
(right). 
The double line stands for the degenerate horizon. 
Only the vicinity of the shaded point at $r=0$ with $t$ kept finite approximates our
original solution. Whereas, the domains bounded by blue shaded lines in each
figure are locally isometric~\cite{Carter}. }
\label{Nariai}
\end{center}
\end{figure}

On the other hand, taking the limit 
$r_{(i)}:=|\,\vect{x}-\,\vect{x}_{(i)}| \to 0$,
we can safely neglect the time-dependence of the metric. 
Hence, the metric at the very neighborhood of each mass point
is approximated by the 
Nariai-Bertotti-Robinson metric~\cite{Nariai,BR} [the direct product of a 2-dimensional anti-de
Sitter (AdS$_2$) and a ($D-2$)-sphere] as 

\begin{align}
\D s^2 _{r_{(i)}\to 0}=-\frac{r_{(i)}^{2}}{\ell_{(i)}^{2}}\D t^2
+\frac{\ell_{(i)}^{2}}{r_{(i)}^{2}}
\D r_{(i)}^{2} + \ell_{(i)}^2 \D \Omega_{D-2}^2
\,,
\end{align}
where 
$
\ell_{(i)}:=[(Q_T^{(i)})^{n_T}(Q_S^{(i)})^{n_S}]^{1/[2(D-2)]}
$
sets the curvature scale
of ${\rm AdS}_2$ and ${\rm S}^2$ at the $i$-th point, and 
$\D \Omega_{D-2}^2$ is the line-element of a $(D-2)$-dimensional
unit sphere.
It has been noticed that the neighborhood of any extremal black
holes can be universally described by the above metric~\cite{Kunduri:2007vf,Astefanesei:2007bf}. 
Figure~\ref{Nariai} compares the geometry of 
the AdS$_2\times {\rm S}^{D-2}$ and that of the extremal RN black hole.

Thus, one anticipates that the metric~(\ref{metric})
describes a system of charged black holes
 with a degenerate event horizon embedded
in the  FLRW universe filled by a fluid~(\ref{EOS}),
which might lead us
to speculate that the causal structure is
found simply by patching
Figures~\ref{fig:FRW} and \ref{Nariai}. 
As pointed out in~\cite{MN}, however, this na\"ive expectation fails
if a single point mass is placed at $r=0$ for $D=4$ and $n_T=1$. 
The solution in~\cite{MOU} has neither a degenerate
event horizon at $r=0$, nor a big-bang singularity at $t=0$.
As is clear from Figure~\ref{Nariai}, the limit $r \to 0$ with $t$ kept
finite does not describe the $(D-1)$-dimensional null surface
(the blue shaded domain corresponding to 
$r\to 0$ with $|t|$ going to infinity), but describes only the point at
the ``throat infinity.''  This lies at the core of why the present
metric admits nonvanishing Hawking temperature.  
In order to elucidate the full causal structure, one has to
discuss these null surfaces with much care, the analysis of which is 
the principal subject of this paper.

A curious property of the present solution~(\ref{metric})
is that the field equations are completely linearized. 
One may encounter the linear gravitational
equations in the context of the solitonic configurations and the supersymmetric
solutions. In the present case, none of these interpretation
fails.
Despite the non-trivial time dependence, 
we are able to superpose the harmonics {\it ad arbitrium}. 
This property might be seemingly contradictory. 
Yet, one faces the same situation  in the Kastor-Traschen solution~\cite{KT}, where  
the black hole collision can be described by multicenter harmonics. 
These spacetimes maintain a special balance 
in close analogy with supersymmetric states. \footnote{
It has been pointed out that in the Kastor-Traschen spacetime 
there exists a super-covariantly
constant spinor by Wick-rotating the coupling constant, $g$,  
in the gauged $N=2$ supergravity as $g\to ig$~\cite{London,KT2}. 
This is not related to the genuine supersymmetry because gauged supergravities
require the negative cosmological constant. Rather,  
it arises from the ``fake'' supergravity
\cite{Behrndt:2003cx,Freedman:2003ax,Grover:2008jr}, 
in which the supercovariant derivative operator is no longer Hermitian, and 
the unwanted ghosts appear due to the non-compact gaugings.  
It is interesting to discuss if the present system can be embedded in the 
fake supergravity. 
}

We will postpone the discussion regarding the multiple black holes in a
separated paper. In this article,  
we shall limit our attention to the simplest case in which 
the harmonics $H_A$'s possess only a monopole term 
$\propto Q_A/r^{D-3}$ at the coordinate center, i.e., the spacetime is spherically symmetric.
In this single-mass case for each harmonic, it is sufficient to consider
 the case of identical charges
$Q_T=Q_S=:Q$,  since $Q_T\ne Q_S$ amounts to a  trivial conformal change 
\begin{align}
g_{\mu \nu }~\to~ (Q_T/Q_S)^{n_T/(D-2)} g_{\mu \nu }
\,,
\end{align} 
with the parameter redefinitions
\begin{align}
 t ~&\to (Q_T/Q_S)^{-n_T/2} \,t\,, \\
t_0&\to ~(Q_T/Q_S)^{1-n_T/2} \,t_0\,. 
\end{align} 
Thus,  we will be concerned with the metric,
\begin{align}
\D s^2= -\Xi^{D-3}\D t^2+\Xi ^{-1}\left( \D r^2+r^2 \D
 \Omega_{D-2}^2\right)\,,
\label{metric_2}
\end{align}
where  
\begin{align}
\Xi =\left( H_T^{n_T}H_S^{n_S}\right)^{-1/(D-2)}\,,
\label{Xi_2}
\end{align}
with
\begin{align}
H_T:&=\frac{t}{t_0}+\frac{Q}{r^{D-3}}\,, \\
H_S:&=1 +\frac{Q}{r^{D-3}}\,.
\end{align}
The constant $Q$ has a physical meaning as a U$(1)$ charge, satisfying
\begin{align}
 \frac{Q}{\sqrt G} =\frac{1}{(D-3)\Omega_{D-2}}\int _S e^{\lambda_A \kappa
 \Phi }F_{\mu \nu }^{(A)}\D S^{\mu \nu }\,,
\end{align} 
where $S$ is an arbitrary closed surface surrounding $r=0$, and    
\begin{align}
\Omega_{D-2}=\frac{2\pi ^{(D-1)/2}}{\Gamma [(D-1)/2]} 
\end{align}
denotes the area of a $(D-2)$-dimensional unit sphere. 
The above first integral is a direct consequence of
Eq.~(\ref{Maxwelleq}). 
Though the charge may take either sign, 
we shall restrict ourselves to the positive case 
($Q>0$) for definiteness of our argument.

\section{Matter fields and singularities}
\label{sec:matter}

Let us first evaluate matter fields and discuss energy conditions. 
Next, we shall discuss the spacetime singularity. 

\subsection{Energy conditions}

Many candidates of  ``black hole solutions'' in 
the expanding universe found in the literature
(Sultana-Dyer solution~\cite{SultanaDyer}, 
McVittie's solution~\cite{Mcvittie1933}, etc)
do not respect energy conditions in the whole of spacetime. 
It is therefore worthwhile to argue whether the 
present spacetime satisfies suitable energy conditions.

Since the explicit metric-form is available, we 
can immediately evaluate the matter contents of 
the solution. Let us consider a (nongeodesic) observer 
whose integral curve is tangent to $\partial /\partial t$. 
The energy densities and the pressures for this observer are given by
\begin{align}
 \rho^{(\Phi)} &=\frac{1}{2}\left[\Xi^{-(D-3)}( \Phi_{,t})^2+\Xi (\Phi_{,r})
 ^2\right]+V \nonumber \\
&=\ma F_1+\ma F_2+\ma F _3>0\,,
\label{rho_Phi}\\
 P_r^{(\Phi)} &=
\frac{1}{2}\left[\Xi^{-(D-3)}( \Phi_{,t})^2+\Xi (\Phi_{,r})
 ^2\right]-V \nonumber \\
&=\ma F _1+\ma F_2-\ma F_3 \,, 
\label{P1}
\\
P_ {\Omega }^{(\Phi)}&=
\frac{1}{2}\left[\Xi^{-(D-3)}( \Phi_{,t})^2-\Xi (\Phi_{,r})
 ^2\right]-V \nonumber \\
&=\ma F_1-\ma F_2-\ma F_3 \,, 
\label{POmega}\\
\rho ^{(\rm em)}&=-P_r^{(\rm em)}=P_{\Omega }^{(\rm em)} 
\nonumber \\
&=\ma F_4>0\,,
\label{rho_em}
\end{align}
where the comma denotes the partial derivative, 
and
\begin{align}
\ma F_1:=& \frac{n_Tn_S(D-3)}{8\kappa^2 (D-2)t_0^2}
\left(\frac{H_S}{H_T}\right)^{n_S(D-3)\over (D-2)}\,,
\label{F1}
\\
\ma F_2:=&\frac{n_Tn_SQ^2(D-3)^3(H_T-H_S)^2 }{8\kappa^2 (D-2) r^{2(D-2)}
H_T^{2+{n_T\over (D-2)}}H_S^{2+{n_S\over (D-2)}}}\,,
\label{F2}
\\
\ma F_3:= &\frac{n_T(n_T-1)}{4\kappa
 ^2t_0^2}\left(\frac{H_S}{H_T}\right)^{n_S(D-3)\over (D-2)}\,,
\label{F3}
\\
\ma F_4:= & \frac{(D-3)^2Q^2
\left(n_TH_S^2+n_S H_T^2\right)}{4\kappa^2 r^{2(D-2)}
H_T^{2+{n_T\over (D-2)}}H_S^{2+{n_S\over (D-2)}}}\,.
\label{F4}
\end{align}
Here, the subscript  $\Omega $ represents any of  
the angular components, 
which are all equal under the
spherical symmetry.
Due to the scalar field, 
the observer sees the nonvanishing energy flux, 
whose radial component
$\ma F_5:=-\Xi ^{-(D-4)/2}T_{tr}$ is found to be
\begin{align}
 \ma F_5= -\frac{n_Tn_SQ (D-3)^2(H_T^{n_T}H_S^{n_S})^{(D-4)\over (D-2)}
(H_T-H_S)}
{4 \kappa ^2(D-2)t_0 r^{D-2}H_T^2H_S} 
\,.
\label{F5}
\end{align} 
From these, one can derive the following useful relations
\begin{align}
&\ma F_1+\ma F_3 =
\frac{n_T^2(D-1)}{8\kappa^2 (D-2)t_0^2}
\left(\frac{H_S}{H_T}\right)^{n_S(D-3)\over (D-2)}\,, \\
&\ma F_4-\ma F_2= \frac{(D-3)^2Q^2H_{ST}^2}
{8\kappa^2 (D-2)r^{2(D-2)}H_T^{2+{n_T\over (D-2)}}H_S^{2+{n_S\over (D-2)}}}\,,
\label{relation_F}\\
&(\ma F_1-\ma F_2)^2 =  (\ma F_1+\ma F_2)^2- \ma F_5^2\,,
\end{align}
where  we have used a shorthand notation
\begin{align}
H_{ST}:=n_SH_T+n_TH_S\,. 
\end{align}
Note that $\ma F_1, \ma F_2 , \ma F_4$, 
$\ma F_1 +\ma F_3$ and $\ma F_4-\ma F_2$
are positive-definite, while $\ma F_3$ is negative for $0<n_T<1$.

At the time $t=t_0$ we have $H_T=H_S$, hence 
the physical quantities evaluated at $t=t_0$ are reminiscent of 
those for the extremal RN solution. We shall refer to the time 
$t=t_0$ as a fiducial time.   At $t=t_0$, 
the energy density of the 
scalar field  is uniform in space as
\begin{align}
\kappa^2\rho ^{(\Phi)}={(D-1)n_T^2 \over 8(D-2)t_0^2}
\label{rho_Phi_0}
\,,
\end{align} 
and that of the electromagnetic field is the
same as the extremal RN solution,
\begin{align}
\kappa^2 \rho ^{(\rm em)}
 =\frac{(D-2)(D-3)Q^2}{2(r^{D-3}+Q)^{2(D-2)/(D-3)}}\,.
\label{rho_em_0}
\end{align}
For $t>t_0$, $\ma F_5$ becomes negative, indicative of the 
scalar field falling into the black hole. As it turns out in the later section, 
this is not the case since the black-hole event horizon is a Killing
horizon. 
Killing horizons represent ``equilibrium states'' of event
horizons, precluding the black hole to grow~\cite{wald2}.
Physically considering, the accretion of the 
scalar field is delicately in balance 
on the horizon with  the 
repulsive forces caused by the U$(1)$ fields.

We are now ready to discuss energy conditions for the 
total energy-momentum tensor. For the observer moving along
$\partial /\partial t$, 
the total energy density and the pressures  are given by
\begin{align}
 \rho &=\rho^{(\Phi)}+\rho ^{(\rm em)}\,, \\
 P_r &=P_r^{(\Phi)}+P_r^{(\rm em)}\,, \\
 P_{\Omega }&=P_{\Omega }^{(\Phi)}+P_{\Omega }^{(\rm em)}.
\label{tot_rhoP}
\end{align}
The simplest and the most intuitive one is the weak energy condition~\cite{HE}, 
which states that the energy density for an 
arbitrary observer with worldline $V^\mu $ is non-negative, 
\begin{align}
T_{\mu \nu }V^\mu V^\nu \ge 0\,. 
\end{align}
Solving the characteristic equation
$\det ({T^\mu }_\nu -\ma T {\delta ^\mu }_\nu )=0$,
we can diagonalize the (total) stress-energy tensor as  
${T^{\hat \mu}}_{\hat \nu}={\rm diag} (-\hat \rho ,
\hat P_r, \hat P_\Omega, ..., 
\hat P_\Omega )$,
where the eigenvalues are given by 
\begin{align}
\hat \rho &=\frac{1}{2}\left[\rho- P_r +\sqrt{(\rho+P_r)^2-4\ma F_5^2
 }\right]\,, 
\\
\hat P_r &=\frac{1}{2}\left[-(\rho- P_r) +\sqrt{(\rho+P_r)^2
-4\ma F_5^2 }\right] \,, 
\\
\hat P_\Omega &=P_\Omega \,. 
\end{align}
In terms of these quantities, the weak energy condition reads~\cite{HE}
\begin{align}
\hat \rho\ge 0\,, \qquad
\hat \rho +\hat P_r\ge 0\,, \qquad
\hat \rho+\hat P_{\Omega } \ge 0\,.
\label{WEC2}
\end{align} 
From  Eqs.~(\ref{relation_F}) and (\ref{tot_rhoP}) 
one obtains 
\begin{align}
[(\rho +P_r)^2-4 \ma F_5^2]^{1/2}=2 |\ma F_1- \ma F_2|\,. 
\end{align}
Noticing $\ma F_4-\ma F_2>0$, it follows that the last two inequalities  
in Eq.~(\ref{WEC2}) hold. Then, 
it remains to examine the positivity of 
$\hat \rho =\ma F_3+\ma F_4 +|\ma F_1-\ma F_2 |$, or
a lack thereof.
In the region $\ma F_1-\ma F_2 \ge 0$, we have
\begin{align}
\hat \rho =(\ma F_1+\ma F_3)+(\ma F_4-\ma F_2)>0 \,,
\label{WECeq0}
\end{align}
 while 
in the region $\ma F_1-\ma F_2 < 0$, we have
\begin{align}
\hat \rho =-\ma F_1+\ma F_2+\ma F_3+\ma F_4 \,.
\end{align}
At first sight, the sign of $\ma F_3+\ma F_4$ might be obscure
in the latter case.   
Yet, using the condition $\ma F_1-\ma F_2 < 0$, 
it turns out to be bounded below as 
\begin{widetext}
\begin{align}
\ma F_3+\ma F_4 &>\left[\frac{n_TH_S^2+n_SH_T^2+n_T(n_T-1)(H_T-H_S)^2}
{4\kappa^2
 t_0^2}\right]\left(\frac{H_S}{H_T}\right)^{n_S(D-3)\over (D-2)}\,,
\label{WECeq1}\\
&=\left[\frac{
(n_S-n_T)H_T^2 +2n_TH_TH_S+n_T^2 (H_T-H_S)^2
}
{4\kappa^2 t_0^2}\right]\left(\frac{H_S}{H_T}\right)^{n_S(D-3)\over (D-2)}\,.
\label{WECeq2}
\end{align}
\end{widetext}
From Eq.~(\ref{WECeq1}) the positivity of $\hat \rho $ 
immediately follows for $n_T\ge 1$. 
In the case of $n_T<1$ [$n_S>(D-1)/(D-3)$], observing 
the inequality 
\begin{align}
 n_S-n_T >\frac{D-1}{D-3}-1=\frac{2}{D-3}>0\,,
\label{WECeq3}
\end{align}
we are led to the conclusion $\hat \rho >0$
from Eq.~(\ref{WECeq2}). 
Hence, the weak energy condition is satisfied in either case.

One can evaluate other types of energy conditions in a similar fashion. 
In the background FLRW universe, 
the acceleration of the  universe is driven when 
timelike convergence condition ceases to be valid~\cite{HE}. 
In general relativity this condition 
is equivalent to saying that the strong energy condition
\begin{align}
\left(T_{\mu \nu }-\frac{1}{D-2}{T^\rho }_\rho g_{\mu \nu }
\right)V^\mu V^\nu \ge 0\,,
\label{SEC}
\end{align}
is violated, 
indicative of peculiar matter fields with large negative pressures.
In terms of eigenvalues of the stress-energy tensor, 
the strong energy condition amounts to  
\begin{align}
&\hat \rho +\hat P_r \ge 0\,, \qquad 
\hat \rho+\hat P_\Omega \ge 0\,, \nonumber \\ 
&(D-3) \hat \rho +\hat P_r+(D-2) \hat P_{\Omega } \ge 0\,.
\label{SEC2}
\end{align}
The last condition is rewritten as
\begin{align}
&(D-3) \hat \rho +\hat P_r+(D-2) \hat P_{\Omega }
\nonumber \\
&~ =(D-2)\left(|\ma F_1-\ma F_2| +\ma F_1-\ma F_2 \right)
\nonumber \\
&~~~~+2 \left[-\ma F_3+(D-3)\ma F_4\right]\,.
\label{SEC3}
\end{align}
In the case of $\ma F_1-\ma F_2 >0$, a straightforward calculation
reduces the right-hand side of the above equation to  
\begin{widetext}
\begin{align}
&8t_0^2\kappa^2 H_T^{2+{n_T\over (D-2)}}H_S^{2+{n_S\over (D-2)}}
\left[(D-2)(\ma F_1-\ma F_2)-\ma F_3+(D-3)\ma F_4 \right]
\nonumber \\
&~~=
\frac{(D-3)^3Q^2t_0^2}{r^{2(D-2)}}
\left\{2(n_TH_S^2+n_SH_T^2)-n_Tn_S(H_T-H_S)^2 \right\} -(D-1)n_T (n_T-2)
H_T^{n_T}H_S^{n_S+2}\,.
\label{SEC4}
\end{align}
\end{widetext}
Since the last term on the right-hand side of 
Eq.~(\ref{SEC4}) is always dominant for $r\to \infty $, 
the strong energy condition violates 
if $n_T>2$, i.e., when the background FLRW universe 
is accelerating $(p>1)$.
While, for $n_T\leq 2$, i.e., when the background FLRW universe 
is decelerating $(p\leq 1)$, we can conclude that the strong energy condition 
holds,  following the same line of argument as 
Eqs. (\ref{WECeq0})--(\ref{WECeq3}) (note that we must also evaluate the case of $\ma F_1-\ma F_2<0$).
Namely, {\it the present spacetime satisfies the strong energy condition if and only if the 
background FLRW universe does. 
}

As shown above, 
 it is dependent on the parameters and restricted to 
the particular spacetime regions whether 
the strong (and the dominant) energy condition holds.  
The failure of these energy conditions is, however, not so serious. 
It is widely accepted that the minimal requirement 
to avoid any pathologies is the null energy condition:
$\hat \rho +\hat P_r\ge 0$ and $\hat \rho+\hat P_{\Omega } \ge 0$. 
If the null energy condition is not satisfied,
undesirable instabilities would develop~\cite{NEC}. 
Since the present system satisfies the weak energy condition, 
the null energy condition automatically holds~\cite{HE}.  
Hence, the present spacetime is free from such annoying plagues.  
This motivates us to study the system as a physically 
acceptable one.

\subsection{Misner-Sharp energy}

A celebrated quantity that characterizes the matter fields in   
spherical symmetry is the Misner-Sharp quasilocal energy~\cite{MS1964}. 
The advantage of the use of Misner-Sharp energy is that 
it also characterizes the  local spacetime  
structure~\cite{Hayward1994,hideki,nozawa}. 
The quasilocal energy is interpreted as a gravitational energy
contained within a closed surface. Especially when the 
spacetime fails to admit globally conserved charges, the 
quasilocal energy  is a helpful local measure to
identify matter distributions.

A major superiority in spherical symmetry lies in the fact that 
we can define covariantly 
the circumference radius as
\begin{align}
R:=r\Xi^{-1/2} 
\label{R}
\end{align}
Using the above circumference radius, 
the $D$-dimensional Misner-Sharp quasilocal energy,
$m$, is given by~\cite{hideki}
\begin{align}
m=\frac{D-2}{2\kappa ^2} \Omega_{D-2}R^{D-3} \left[1- g^{\mu \nu }
(\nabla_\mu R)(\nabla_\nu R)\right]\,. 
\label{MSmass}
\end{align}
The Misner-Sharp mass is defined uniquely by the  
metric components and its first derivatives, and does not require the
asymptotic conditions of spacetime. Once the spherical surface 
specified by the circumference radius is fixed, the Misner-Sharp mass
is given without ambiguity. Such a quasi-localization is possible
on account of spherical symmetry, in which no gravitational wave is
generated.

\begin{widetext}
The present spacetime metric~(\ref{metric}) yields
\begin{align}
m=&\frac{\Omega_{D-2}|r|^{D-3}}
{8(D-2)t_0^2\kappa^2 r^{2(D-1)}
H_T^{2-[(D-3)n_T]/[2(D-2)]}
H_S^{2-[(D-3)n_S]/[2(D-2)]}}\nonumber \\
& \times 
\left[
n_T^2r^{2D}H_T^{n_T}H_S^{n_S+2}
+4(D-2)(D-3)Qr^{D+1}t_0^2H_TH_SH_{ST}
-(D-3)^2Q^2r^4t_0^2H_{ST}^2
\right]
\,.
\label{MSmass2}
\end{align}
We can provide  each term in Eq.~(\ref{MSmass2}) with physical interpretation  
by evaluating at the fiducial time $t=t_0$, at which 
the Misner-Sharp energy reduces to 
\begin{align}
 m(t_0, R)=&\frac{\Omega_{D-2}n_T^2r^{D-1}H_S^{(D-1)/(D-3)}}
{8(D-2)\kappa^2 t_0^2}+\frac{D-2}{2\kappa^2 H_S}\Omega_{D-2}Q
+\frac{D-2}{2\kappa^2 }\Omega_{D-2}Q\nonumber \\
=&
\frac{\Omega_{D-2}}{D-1}R^{D-1}\rho ^{(\Phi)}(t_0)+\Omega_{D-2}
\int^R_{Q^{1/(D-3)}}\D R R^{D-2}\rho ^{(\rm em)}(t_0,
 R)+\frac{(D-2)\Omega_{D-2}}{2\kappa^2 }Q
\,.
\label{MSmass3}
\end{align}
\end{widetext}
The first term in the above equation corresponds to the energy of 
the scalar field, and the last two terms to the U$(1)$ energies
outside and inside the black hole.

For $Q=0$, only the first term in Eq.~(\ref{MSmass2}) survives, yielding
\begin{align}
 m&= \frac{(D-2)\Omega_{D-2}a^{D-3}r^{D-1}}{2\kappa ^2}\left(
\frac{\D a}{\D \bar t}\right)^2 
\nonumber \\
&=\frac{\Omega_{D-2}}{D-1}(ar)^{D-1}\rho ^{(\Phi)}\,,
\end{align}
as expected for the background FLRW universe. 
This implies also that the first term in Eq.~(\ref{MSmass2})
is the contribution of a scalar field.

\subsection{Curvature singularities}

We can see easily that 
the dilaton profile~(\ref{Phi}) and the 
square of the electromagnetic fields
\begin{widetext}
\begin{align}
F^{(T)}_{~\mu \nu }F^{(T)\mu \nu }&=-\frac{4\pi (D-3)^2 Q^2}{r^{2(D-2)}}
\left(\frac{H_S^{n_S(D-4)}}
{H_T^{(4-n_T)D+4(n_T-2)}}\right)^{1/(D-2)}\,,
\\
F^{(S)}_{~\mu \nu }F^{(S)\mu \nu }&=-\frac{4\pi (D-3)^2Q^2}{r^{2(D-2)}}
\left(\frac{H_T^{n_T(D-4)}}
{H_S^{(4-n_S)D+4(n_S-2)}}\right)^{1/ (D-2)}\,,
\end{align} 
diverge at  
\begin{align}
t= t_s(r):=-Q/r^{D-3}, ~~{\rm and}~~~ 
r^{D-3}=-Q \,.
\label{singualarity}
\end{align} 
One may thence deduce that these surfaces correspond to the 
genuine spacetime singularities characterized by the 
divergence of scalar quantities of curvature.   
As a simple illustration, 
let us see the square of the Weyl-tensor 
\begin{align}
\ma  C_{\mu \nu \rho \sigma }
\ma  C^{\mu \nu \rho\sigma }={(D-3)\over (D-1)}\,\ma  W^2 \,,
\end{align} 
where (see e.g., appendix of~\cite{hideki} for derivation)
\begin{align}
\ma  W= -\frac{(D-3) Q
\left[(D-3)Qr^3 \left\{
H_{ST}^2+2(n_SH_T^2+n_TH_S^2)
\right\}-2(D-1)r^D H_TH_SH_{ST}\right]
}
{2 r^{2D-1}H_T^{2+n_T/(D-2)}H_S^{2+n_S/(D-2)}}
 \,.
\end{align}  
\end{widetext}

One finds that the  the above curvature invariant quantity necessarily 
blows up at~(\ref{singualarity}), where 
the inverse of the lapse function vanishes. 
Since the circumference radius vanishes
at~(\ref{singualarity}), these two singularities 
are both sitting at the physical center $R=0$.  
The other  curvature scalar  quantities 
(e.g. 
$\ma R_{abcd}\ma R^{abcd}$ and  $\ma R$)
do indeed diverge at these central points.

It is notable to remark 
that the $t=0$ surface is completely regular, since 
the curvature invariants are bounded therein. 
The nonzero U$(1)$ charge smoothenes the big-bang singularity which 
exists in the background expanding universe (see Figure~\ref{fig:FRW}). 
Note also that the  null surface $r=0$ is not singular, as in the 
 case of~\cite{MN}. We may therefore extend the spacetime
across the $r^{D-3}=0$ surface, i.e., the negative $r$ region for even
dimensions, and the  $r^2<0$ region for odd dimensions.

Accordingly,  
the allowed coordinate regions are 
$H_T^{n_T}H_S^{n_S}>0$,  
which is arranged to give 
\begin{align}
t> t_s(r) ~~{\rm with}~~~ r^{D-3}>-Q
\end{align}
 or 
\begin{align}
t < t_s(r) ~~{\rm with}~~~ r^{D-3}<-Q
\,. 
\end{align}
We are not interested in the latter domain since it turns out to be disconnected 
with the outside region $r>0$ by the singularity. 
For this reason, we shall focus on the coordinate ranges
\begin{align}
 t> t_s(r)\,, \qquad
r^{D-3}>-Q\,, 
\end{align}

Next, it is important to address the causal structure of these singularities.
We can intuitively expect that both of these singularities have the timelike structure, since the 
Misner-Sharp mass is unbounded below as approaching these singularities.
To establish the timelike signature of singularities, 
it is sufficient to show that there
exist an infinite number of radial null geodesics stemming 
from and terminating into the singularities~\cite{MN,NM}.

In the vicinity of the singularity $r=-Q^{1/(D-3)}$, let us assume the 
following asymptotic form of the radial null geodesics, 
\begin{align}
  r+Q^{1/(D-3)} =C_0 (t-t_1)^q\,,
\end{align} 
where $q(>0)$ and $C_0$ are constants. 
Substituting this ansatz into 
the outgoing radial null condition 
$\D r /\D t=\Xi ^{(D-2)/2}$,  we obtain 
\begin{align}
q&=\frac{2}{n_S+2}\,, 
\nonumber \\
C_0^{1+n_S/2}&=\frac{n_S+2}{2} Q^{n_S/2}
 \left(1-\frac{t_1}{t_0}\right)^{-n_T/2}\,. 
\end{align} 
This verifies that infinite geodesics 
parametrized by the departure time $t_1 (<t_0)$
emanate from the singularity. 
For the ingoing null geodesics, we can also 
arrive at the conclusion that there exist infinite ingoing null
geodesics running into the singularity.   
The proof proceeds similarly for the singularity 
$t= t_s(r)$: 
it is shown that there exist infinite ingoing and outgoing geodesics
falling into and emerging from the singularity at the time $t>t_0$
for $r<0$ and at the time $t<0$ for $r>0$.  
It follows that both of these two singularities have a timelike structure, i.e., 
these are locally naked singularities.  
As argued in~\cite{HH}, these timelike singularities 
are associated with the U$(1)$ charge, not with the background universe.

In order to verify whether these singularities are visible to observer
far in the distance from the central region, 
we have to trace the geodesics all the way out to infinity. 
We will numerically confirm in later section that the 
singularities $t_s (r)$ and  $r=-Q^{1/(D-3)}$ lying in the $r<0$ region 
are contained inside the black hole. 
On the other hand, we will see that the singularity $ t_s (r)$ 
emerging in the $r>0$ region is not covered by the event horizon.

\section{Spacetime structures}
\label{sec:spacetime}

We shall explore the spacetime structure of the ``black hole
candidate'' (\ref{metric}). 
In order to conclude that the present spacetime is indeed qualified as
 a black hole,  
we need to appreciate the infinite future of the spacetime
to assess  the locus of  the event horizon. 
Thus, 
the notion of event horizon is of less practical use
especially in the time-evolving spacetimes.  
In the dynamical spacetime where the null geodesic motions are not
solved analytically, it is much more advantageous to focus on the 
{\it trapped surface}, which 
is locally defined  and hence free from 
such a difficulty. When  
a catastrophic gravitational collapse occurs to form a black
hole,  there arises a region where even ``outgoing'' null rays are
dragged  back due to strong gravity of the black hole. 
These null rays with negative expansion will be absorbed into the 
hole, characterizing the intuitive idea of the black hole as a region of no escape. 
For each time slice, 
this surface defines an {\it apparent horizon}~\cite{HE} as an outermost 
boundary of the trapped region 
in the asymptotically flat spacetimes. 
Hayward generalized these quasilocal notions to define a class of 
{\it trapping horizons}~\cite{Hayward1993}.  
Trapping horizons are associated not only with black holes 
but also with white holes and cosmological ones, suitable for the 
present context.

Since the event horizons and trapping horizons are 
conceptually different, 
there is no {\it a priori} relationship between them
(the black hole in stationary spacetime is exceptional, for which the
event horizon and the trapping horizon coincide~\cite{HE,HIW}).  
Nevertheless,  we can rely on the Hawking's theorem that 
all trapped surfaces appear 
inside a black hole 
under the null energy condition 
provided the spacetime is asymptotically flat with some 
additional technical assumptions (see Proposition 9.2.8 in~\cite{HE} for
the proof). Although the present spacetime is not asymptotically flat, 
taking into the fact that the metric is well-behaved 
off of the central singularity at $R=0$, this criterion is of use for our study. 
This expectation will be backed by numerical analysis of null geodesic
motions.

For simplicity of our argument, we shall focus our attention only
to the 4-dimensional case.  
We also assume  $t_0>0$ and $Q>0$.  
We define dimensionless variables $\tilde t:=t/t_0$, 
$\tilde r:=r/Q$
and work in the dimensionless metric 
$Q^{-2} \D s^2=:\D \ti s^2 =\ti g_{\mu \nu }\D \ti x^\mu \D \ti x^\nu $
to simplify our discussion. 
To be specific, we work in the metric
\begin{align}
 \D \ti s^2=-\tau^2 \Xi \D \ti t^2 +\Xi ^{-1}
\left[\D \ti r^2+\ti r^2 \left(\D \theta^2 +\sin^2 \theta \D \phi^2 
\right)\right]\,,
\label{metric0}
\end{align}
where
\begin{align}
\tau:=\frac{t_0}{Q}\,, \qquad 
\Xi = \frac{\ti r^2}{[(\ti t\ti r+1)^{n_T}(\ti r+1)^{n_S}]^{1/2}}\,, 
\end{align}
with
\begin{align}
n_T+n_S=4
\,,
\end{align}
for which the dimensionless circumference radius is defined by
\begin{align}
\ti R = [(\ti t\ti r+1)^{n_T}(\ti r+1)^{n_S}]^{1/4}\,.
\end{align} 
The metric is parametrized by two dimensionless
parameters $\tau $ and $n_T$.
In what follows, 
dimensionless variables associated with the metric~(\ref{metric0})
will be denoted with tilde.

The  physical meaning of $\tau$ is na\"ively given as follows.
The energy density of the scalar field at $t=t_0$  
is $\rho^{(\Phi)}\propto 1/t_0^2$ [Eq.~(\ref{rho_Phi_0})].
While, 
for the extreme RN black hole with the charge $Q$, 
the total energy density of the U(1) fields 
evaluated on the horizon is given by 
$\kappa^2 \rho^{\rm (em)}|_{\scriptscriptstyle H}\propto 1/Q^2$
[Eq.~(\ref{rho_em_0})]. 
Thus for the time-dependent black hole
we can claim that $\tau$ is related to 
the ratio of two energy densities at the event horizon as
\begin{align}
\tau^2\sim
{
\rho^{\rm (em)}
\over 
 \rho^{(\Phi)}
}
\Big{|}_{\scriptscriptstyle H}
\,. 
\end{align}
A more rigorous treatment shown below [see Eq.~(\ref{tau_meaning})] 
confirms that the above estimate is indeed true.

Before going to the detailed analysis, we 
classify our spacetimes into nine types 
[from Case I-(i) to Case III-(iii)]. 
Classification of Cases I--III is dependent on 
the expansion rate of the universe, i.e.,
whether the universe is decelerating
($n_T<2$), marginal ($n_T=2$) or accelerating ($n_T>2$).
The parameter $\tau$ is also important in the subclassification for each
Class I--III, whereby
the global structures of the solution change drastically. 
We summarize our classification in Table~\ref{Table},
which includes some important results given later.

\subsection{Trapping horizons}

A trapped surface is defined on a closed and orientable spacelike 
surface $S$. Since the present spacetime admits an SO$(3)$-symmetry, 
we take $S$ as a metric sphere. 
Let us introduce the Newman-Penrose null tetrads as
\begin{align}
l_\mu\D \tilde x^\mu&=\sqrt{\frac{\Xi}{2}}(-\tau 
\D \tilde t+\Xi^{-1}\D \tilde r)\,, 
\nonumber \\
 n_\mu\D \tilde  x^\mu
&=
\sqrt{\frac{\Xi}{2}}
(-\tau \D \tilde t-\Xi^{-1}\D \tilde r)\,,
\label{NPtetrads} \\
 m_\mu\D \tilde  x^\mu &=\frac{\tilde  r}{\sqrt{2\Xi}}
(\D \theta + i \sin \theta \D \phi )\,,\nonumber 
\end{align}
with $\bar m_\mu $ being a complex conjugate of $m_\mu $. 
These null vectors satisfy $l^\mu n_\mu=-1=-m^\mu\bar m_\mu$. 
$l^\mu $ and $n^\mu $ are the future-directed radial null vectors. 
Using these tetrads, 
expansions are defined by 
\begin{align}
\ti \theta_+:=2m^{(\mu }\bar m^{\nu )}\ti \nabla_\mu l_\nu \,, \qquad
\ti \theta_-:=2m^{(\mu }\bar m^{\nu )}\ti \nabla_\mu n_\nu \,,
\end{align}
where the operation of ``{\rm Re}'' has been omitted because  
the vanishing rotation is obvious. 
The present metric~(\ref{metric0}) computes to give
\begin{align}
 \ti \theta_\pm =\frac{
n_T\ti r^2H_S\sqrt{H_T^{n_T}H_S^{n_S}}\pm \tau \left(
4\ti rH_TH_S-H_{ST}\right)
}{2\sqrt 2 \tau \ti r^2H_T^{1+n_T/4}H_S^{1+n_S/4}}\,.
\end{align}
These expansions characterize the rate of divergence 
of outgoing and ingoing light rays. 
It should be remarked, however, that values of expansions themselves are not
invariant quantities due to the relative normalization of 
$l^\mu$ and $n^\mu $. Only their product 
\begin{align}
 \ti \theta_+\ti \theta_-=-2\ti R^{-2}(\ti \nabla_\mu \ti R)
(\ti \nabla^\mu  \ti R)
\label{thp_them}
\end{align}
has a covariant meaning.

A metric sphere is said to be {\it trapped} ({\it untrapped}) 
if $\ti \theta_+\ti \theta_- > 0$ ($\ti \theta_+\ti \theta_- < 0$), 
and {\it marginal} when $\ti \theta _+\ti \theta_- =0$. 
A {trapping horizon} is the closure of a hypersurface foliated by
marginal surfaces, i.e., 
trapping horizons occur at $\ti \theta_\pm =0$.

It is instructive first to see the case of FLRW universe ($Q=0$), 
in which only $\theta _-=0$ has a solution
\begin{align}
r_{\rm TH}={4t_0 \over n_T}\, \left(\frac{t}{t_0}\right)^{\,1-{n_T/2}}\,.
\end{align} 
Using the relations~(\ref{tbar}) and (\ref{scale_factor}),
the trapping horizon is expressed in terms of the 
cosmic time $\bar t$ as 
$r_{\rm TH}=p^{-1}\bar t_0 (\bar t/\bar t_0)^{1-p}$. 
Written in the  the circumference radius $R=ar$,
this is translated into  
\begin{align}
 R_{\rm TH }=a \left(\frac{
\D a}{\D \bar t}\right)^{-1}\,,
\end{align}
 which is the 
Hubble horizon as desired.  
This kind of trapping horizon is associated to the background universe--the 
{\it past trapping horizon} in the terminology of
Hayward~\cite{Hayward1994}--manifesting a situation that 
even the ingoing null rays are pushed outwards
due to the cosmic expansion.

Unfortunately, for $Q\ne 0$ it is not feasible to obtain the 
orbits of trapping horizons in the explicit analytic form 
$\ti t=\ti t_{\rm TH}(\ti r)$ 
unless $n_T$ takes an integer. 
Reminding that the trapping horizons
occur where $\ti \nabla _\mu \ti R $ becomes null, we find that
it is more helpful to express the positions of trapping horizons
in terms of the circumference radius $\ti R$.    
Now, $\ti t$ is expressed by $\ti R$ and $\ti r$ as 
\begin{align}
 \tilde t=\frac{1}{\tilde r} \left\{ \left[\frac{\tilde
 R^4 }{(\ti r+1)^{n_S}}\right]^{1/n_T}-1\right\}\,. 
\label{t_Rr}
\end{align}
Inserting this relation into Eq.~(\ref{metric}), 
we can write the metric in terms of $\ti r$ and $\ti R$ as 
\begin{align}
 \D \ti s^2 =&-\tau^2 \biggl[
\frac{\D
 \ti r}{\ti r\ti R}\left\{1-\frac{\ti R^{4/n_T}(n_T+4\ti r)}{n_T(\ti
 r+1)^{4/n_T}}\right\}
\nonumber \\
&+\frac{4\ti R^{n_S/n_T}\D \ti R}{n_T\ti R (\ti r+1)^{n_S/n_T}}
\biggr]^2+\frac{\ti R^2}{\ti r^2}\D \ti r^2 +\ti R^2\D \Omega_2^2\,. 
\label{metricRr}
\end{align}
Setting $\ti R={\rm constant}$, 
one obtains the induced metric of the constant $\tilde R$ surfaces,  
\begin{widetext}
\begin{center}
\begin{table}[t]
\caption{Classification of the spacetime structures of the 
solution~(\ref{metric0}). The parameter $n_T$ corresponds to 
the steepness parameter of the dilaton potential, which is related to the 
expansion exponent of  the FLRW universe $a\propto \bar t^p$ as
Eq.~(\ref{scale_factor}), while  
$\tau :=t_0/Q$ measures the ``nonextremality'' of the black hole. 
This classification is based on the spacetime structures
discussed in the text.
In the last column, ``BH'' and ``NS'' represent whether the 
spacetime describes a 
regular black hole or a naked singularity without an event horizon.}
\begin{tabular}{c|c||c|c||c|c|c|cl} 
\hline\hline
\multicolumn{2}{c||}{\raisebox{-.5em}{Case}}
 & \raisebox{-.5em}{$n_T$} 
 & \raisebox{-.5em}{$\tau $}
 &number of
 &\multicolumn{4}{c}{causal structure}
\\[-.5em] \cline{6-9}  
\multicolumn{2}{c||}{~}  &   & & horizons
&infinity & near-horizon  & \multicolumn{2}{c}{global structure} 
\\ \hline\hline

\multicolumn{2}{c||}{extreme RN}
& $n_T=0$     & any          & 1               
& Minkowski  &Fig.~\ref{Nariai} (left)& Fig.~\ref{Nariai} (right)&:\,BH
\\ \hline
 
\raisebox{-.5em}{ I} &{ (i)}   &$0<n_T< 4/3$   & & 
& Fig.~\ref{fig:FRW}(1) &   &  &
 \\[-.3em] \cline{2-3}\cline{6-6}

\raisebox{-.5em}{(decelerating universe: $p<1$)}
&{ (ii)}  &$n_T=4/3$  & any  & 2  
& Fig.~\ref{fig:FRW}(2) & Fig.~\ref{PDstatic}(A) 
& Fig.~\ref{PD}(a)&:\,BH
\\[-.5em]  \cline{2-3}\cline{6-6}

&{ (iii)}   &$4/3<n_T< 2$  &   & 
& Fig.~\ref{fig:FRW}(3) &   &  
&
\\ \hline

\raisebox{-.5em}{ II}  & {(i)}  &
                     & $\tau>1 $    & 2 &  
           & Fig.~\ref{PDstatic}(A) &  Fig.~\ref{PD}(a)&:\,BH
\\[-.2em] 
\cline{2-2} \cline{4-5} \cline{7-9}

\raisebox{-.5em}{(Milne universe: $p=1$)} & {(ii)}
 & $n_T=2$     & $\tau=1$     & \raisebox{-.5em}{1}  
& Fig.~\ref{fig:FRW}(4)       & Fig.~\ref{PDstatic}(B) & 
\raisebox{-.5em}{Fig.~\ref{PD}(b)}&\raisebox{-.5em}{:\,NS}
\\[-.5em]
 \cline{2-2} \cline{4-4} \cline{7-7}

&{(iii)}  &                     & $\tau < 1$   &   
&     &        Fig.~\ref{PDstatic}(C) & &
 \\ \hline

\raisebox{-.5em}{ III} &{ (i)}
              &          & $\tau>\tau_{\rm cr}$& 3 
&            & Fig.~\ref{PDstatic}(D) &  Fig.~\ref{PD}(d)&:\,BH
\\[-.2em] 
 \cline{2-2} \cline{4-5} \cline{7-9}

\raisebox{-.5em}{(accelerating universe: $p>1$)} 
 &{ (ii)}  &$2<n_T<4 $     & $\tau=\tau_{\rm cr}$& 2  
&  Fig.~\ref{fig:FRW}(5)       & Fig.~\ref{PDstatic}(E) &  
Fig.~\ref{PD}(e)&:\,BH  
\\[-.4em]  \cline{2-2} \cline{4-5} \cline{7-9}

&{(iii)} &                    & $\tau<\tau_{\rm cr}$& 1   
&            & Fig.~\ref{PDstatic}(C) &  Fig.~\ref{PD}(c)&:\,NS
\\ \hline

\multicolumn{2}{c||}{RNdS}
     & $n_T=4$   & any          & 3   
&  dS        & n/a & Fig.~\ref{PDstatic}(D)&:\,BH
\\ 
\hline \hline

\end{tabular}
\label{Table}
\end{table}
\end{center}
\end{widetext}
\begin{align}
\D \ti s^2_{R}
=-\frac{1}{\ti R^2 \ti r^2}\left[\tau^2 \left\{
\frac{\ti R^{4/n_T}(n_T+4\ti r)}{n_T(1+\ti r)^{4/n_T}}-1
\right\}^2-\ti R^4\right]\D \ti r^2\,,
\label{Rconst}
\end{align}
where we have discarded the angular parts which are 
irrelevant to the discussion of trapping horizon.

Figure~\ref{fig:Rr} depicts the $\ti R-\ti r$ diagram, from which
we can read off the signature of $\ti R={\rm constant}$ surfaces.  
From Eq.~(\ref{thp_them}), the region where the 
$\ti R={\rm constant}$ surface is timelike
(spacelike) consists of the untrapped (trapped) surfaces.
The thick lines in Figure~\ref{fig:Rr} designate the
loci of trapping horizons at which the terms in square-bracket of 
Eq.~(\ref{Rconst}) vanish, $\D \ti s^2_{R}=0$. 
The spacetime has at most three types of 
trapping horizons, 
$\ti R_2(\ti r)\le \ti R_{1,-}(\ti r) \le \ti R_{1,+}(\ti r)$,
 where 
$\ti R_2(\ti r)$ satisfies  
\begin{align}
\tau \left[
\frac{\ti R_2(\ti r)^{4/n_T}(n_T+4\ti r )}{n_T(1+\ti r)^{4/n_T}}-1
\right]+\ti R_2 (\ti r)^2=0\,,
\label{THeq1}
\end{align}
whereas $\ti R_{1,-}(\ti r)$ and $\ti R_{1,+}(\ti r)$
are two positive roots (if they exist) of the equation  
\begin{align}
\tau \left[
\frac{\ti R_1 (\ti r)^{4/n_T}(n_T+4\ti r )}{n_T(1+\ti r)^{4/n_T}}-1
\right]-\ti R_1 (\ti r)^2=0\,.
\label{THeq2}
\end{align}
Observe that the definition of $\ti R_1(\ti r)$ and 
$\ti R_2(\ti r)$ differs from that in the previous paper~\cite{MN},
where $\ti R_{1}(\ti r)$ and $\ti R_{2}(\ti r)$ were 
defined by the surfaces of $\ti \theta_-=0 $ and  
$\ti \theta_+=0$, respectively. In the $\ti r<0$ region, 
the surfaces $\ti R_1 (\ti r) $ and $\ti R_2(\ti r)$ defined by 
Eqs.~(\ref{THeq2}) and (\ref{THeq1}) are 
not always coincident with the $\ti \theta_-=0$ and 
$\ti \theta_+=0$ surfaces.    
Still, it is found that in the $\ti r>0$ region 
$\ti \theta_-=0$ holds at $\ti R_{1}(\ti r)$ and 
$\ti \theta_+=0$ holds at $\ti R_2(\ti r)$.

\begin{widetext}

\begin{center}
\begin{figure}[h]
\includegraphics[width=16cm]{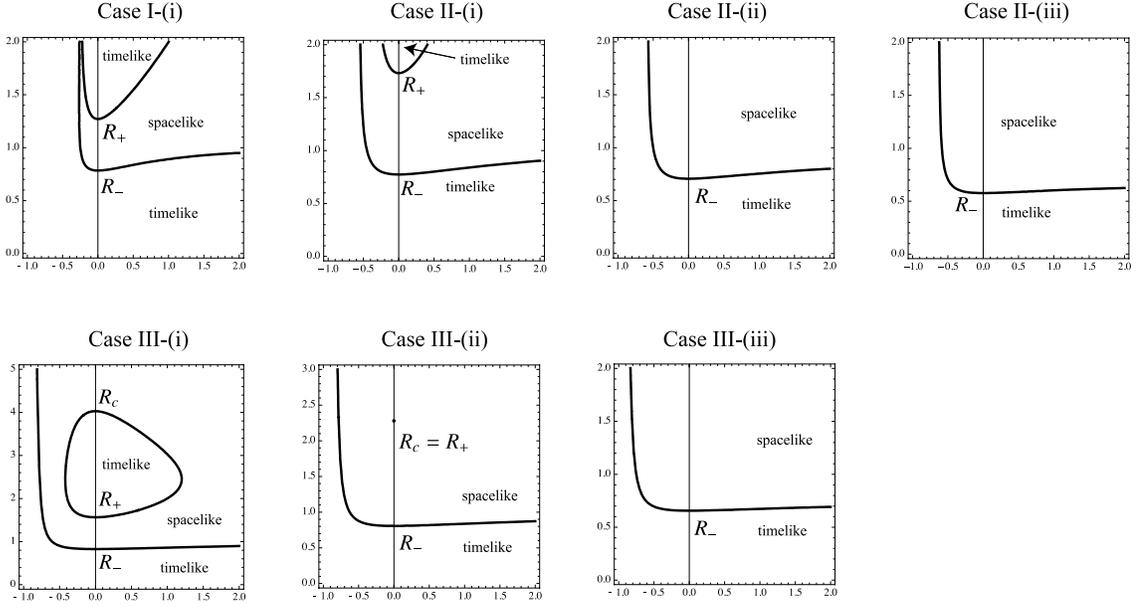}
\caption[width=16cm]{
{Typical plots of 
circumference radius $\ti R$ (vertical axis)
against $\ti r$ (horizontal axis) for
Case I-(i) ($n_T=1$ with $\tau=1$), 
Case II-(i) ($n_T=2$ with $\tau=3/2$), 
Case II-(ii) ($n_T=2$ with $\tau=1$),
Case II-(iii) ($n_T=2$ with $\tau=1/2$), 
Case III-(i) ($n_T=3$ with $\tau=3$), 
Case III-(ii) ($n_T=3$ with $\tau=\tau_{\rm cr}$) and 
Case III-(iii) ($n_T=3$ with $\tau=1$).
The curve of Case I-(i) is the representative for Case I.
The thick lines denote
trapping horizons, across which the $\ti R={\rm constant}$ surface
changes the signature. The region ``spacelike'' 
(and respectively ``timelike'') 
stands for the domain where the
$\ti R={\rm constant}$ surfaces are spacelike (timelike). 
In Case III-(i), the bounded region tends to shrink as $\tau $ decreases
and eventually degenerate into a point shown in Case III-(ii).
}}
\label{fig:Rr}
\end{figure}
\end{center}
\end{widetext}

In Figure~\ref{fig:Rr} we display seven plots
of trapping horizons according to the classification shown in
Table~\ref{Table}.
[We have dropped Cases I-(ii) and I-(iii) because these plots are the same as that for Case I-(i).]
Our previous paper~\cite{MN} discussed the spacetime with  
$n_T=1$  belonging to Case I-(i), where two trapping horizons
$\ti R_{1,-}(\ti r)$ and $\ti R_{2}(\ti r)$ arise. 
Case II-(i) behaves similarly as Case I. 
The universe in Case II undergoes a marginal acceleration 
$a\propto \bar t$.
Our main interest is toward Case III,  
where the background FLRW universe is accelerating. 
Figure~\ref{fig:Rr} illustrates  
that the region $\ti R>\ti R_{1,+}(\ti r)$ 
with $\ti r>0$ is composed of trapped
surfaces, in accordance with the acceleration of the background FLRW
universe.

Let us uncover the properties of these trapping horizons. 
Looking at Eq.~(\ref{THeq1}), 
one finds that the trapping horizon 
$\ti R_2(\ti r)$ approaches to the constant  $\ti R_2(\ti r) \to \sqrt{\tau }$ 
in the limit of $\ti r\to \infty $.   
Noting the inequality  $\D \ti R_2(\ti r)/\D \ti r>0$
for $\ti r> 0$, $\ti R_2 (\ti r)\le \sqrt{\tau }$ is concluded in the
$\ti r>0 $ region. 
In Case I ($n_T<2$),  $\D \ti R/\D \ti r< (>) 0$ holds for 
$\ti R_2< (>)\ti R_{2, <} $ 
in the region $\ti r<0$, where 
\begin{align}
\ti R_{2,<}:=\sqrt{\frac{2\tau }{2-n_T}}\,.
\end{align}
While, for $n_T\ge 2$, $\D \ti R_2/\D \ti r<0$ holds 
in the $\ti r<0$ region.
Thus, in every case the trapping horizon 
$\ti R_2(\ti r)$ develops and diverges as
$\ti r\to -n_T/4$.

It is easy to find that $\ti R_{1,+}(\ti r)$ is 
confluent with $\ti R_{1,-}(\ti r)$ at the constant radius
$\ti R=\ti R_{1,>}$, where
\begin{align}
 \ti R_{1,>} :=  \sqrt{\frac{2\tau}{n_T-2}}\,.
\label{Rc=Rp}
\end{align}
at which $\ti R_{1,\pm}(\ti r)$
has a vertical tangent. Hence, the largest trapping horizon 
$\ti R_{1,+}(\ti r)$ exists if and only if $n_T>2$. 
For $n_T < 2$ and $n_T=2$ with $\tau >1$, 
the trapping horizon $\ti R_{1,-}(\ti r)$ diverges 
as $r\to-n_T/4$.  For $n_T>2$, $\D \ti R_{1, -}/\D \ti r > (<)0$
and $\D \ti R_{1, +}/\D \ti r < (>)0$ hold 
in the region $\ti r>(<)0$. 
Denoting $\ti R_{-}:= \ti R_2(0)$, 
$\ti R_{+}:= \ti R_{1,-}(0)$ and 
$\ti R_c:= \ti R_{1,+}(0)$, it then follows that 
$\ti R_+<\ti R_{1,-}(\ti r)<\ti R_{1,+}(\ti r)<\ti R_c$
for Case III-(i).

The signature of trapping horizons is an important issue to be discussed. 
Differentiating Eq.~(\ref{THeq1}),  
we obtain the expression of $\D \ti R_2/\D \ti r$ along the trapping horizon. 
Substituting this into the metric~(\ref{metricRr}), 
we can find the induced metric on the trapping
horizon.  After a simple but tedious algebra,  we obtain 
\begin{align}
\D \ti s^2 _{\rm TH}=-\frac{4(1+\ti r)^2}{n_S\ti r^2 \ti R_2^4}
\biggl[&
n_T\ti R_2^4+2 \ti R_2^2 (\tau-\ti R_2^2)\nonumber \\
& +\frac{4n_S\ti r^2(\tau -\ti R_2^2)^2}{(n_T+4\ti r)^2}\biggr]\D \ti R_2^2\,.
\label{metric_TH2}
\end{align}
It immediately follows that the cases of $n_T\ge 2$ and 
$n_T<2$ with $\ti R<\ti R_{2,<}$  assure 
that the trapping horizon
$\ti R_2(\ti r)$  is always timelike $\D \ti s_{\rm TH}^2<0$. 
We can numerically check that $\ti R_{2,-}$ is timelike
also for  $\ti R> \ti R_{2,<}$.

Similarly, along the trapping horizon $\ti R_1$, we have
\begin{align}
\D \ti s^2 _{\rm TH}= -
\frac{4(1+\ti r)^2}{n_S\ti r^2 \ti R_1^4}
\biggl[&
n_T\ti R_1^4-2 \ti R_1^2 (\tau+\ti R_1^2)\nonumber \\
& +\frac{4n_S\ti r^2(\tau +\ti R_1^2)^2}{(n_T+4\ti r)^2}\biggr]\D \ti R_1^2\,.
\label{metric_TH1}
\end{align}
Inspecting Eq.~(\ref{Rc=Rp}), the trapping horizon $\ti R_{1,+}$
is necessarily timelike. On the other hand, 
to see the signature of $\ti R_{1,-}$,  
numerical calculation is necessary. Nevertheless, we can see the universal
behavior around $\ti r= 0$, where the last term in the square bracket of Eq.~(\ref{metric_TH1})
is negligible. Taking into account Eq.~(\ref{Rc=Rp}), 
it is found that $\ti R_{1,-}$ is spacelike near $\ti r=0$.

An analytic estimation is possible 
for the background FLRW universe $Q=0$, 
for which Eq.~(\ref{metric_TH1}) reduces to
\begin{align}
 \D s^2_{\rm TH} =\frac{4-3n_T }{n_S}
\D R_1^2 \,,
\end{align}
implying that the signature of trapping horizon is timelike for 
$p>1/2$ ($n_T>4/3$), spacelike for $p<1/2$ ($n_T<4/3$) 
and null for $p=1/2$ ($n_T=4/3$). 
This accounts for the subclassification of Case I shown in Fig.~\ref{fig:FRW}.
The behaviors of trapping horizon $\ti R_{1,-}$ in the asymptotic region
are analogous to those in the FLRW case.

\subsection{Event horizons}

In order to identify the locus of event horizon,
we follow the strategy laid out in our previous paper~\cite{MN}. 
We shall first discuss the geometry of the near-horizon metric and 
demonstrate that the null surfaces of ``event-horizon candidate'' in our dynamical
spacetime~(\ref{metric0}) are described by the Killing horizons.  
Our next tactics  is to confirm these local horizons are 
indeed qualified as true horizons in the original spacetime by
computing the null geodesic motions numerically. 
These logical steps  will guide us in the right direction 
for obtaining the global causal structure.

\subsubsection{Near-horizon geometry}

The analysis in Sec.~\ref{sec:preliminaries}
reveals that the limit $\ti r \to 0$ and 
$\ti t$ being finite corresponds to the 
``throat'' geometry. This fact leads us to speculate that 
the event horizons in the present spacetime, if exist, 
correspond to the null surfaces at $\ti r \to 0$ with $\ti t \to \pm \infty $, i.e., 
the infinite redshift and blueshift surfaces 
``joined'' at the throat. 
To discuss these null surfaces in more detail, 
it is useful to take the {\it 
near-horizon limit}, defined by 
\begin{align}
\tilde t ~\to ~\frac{\tilde t}{\epsilon }\,, 
\qquad 
\tilde r ~\to ~\epsilon \tilde r\,, 
\qquad 
\epsilon~\to ~0\,.
\label{NHlimit}
\end{align}
After the rescalings, the metric is free from $\epsilon $,  and 
we can obtain the near-horizon geometry 
\begin{align}
\D \tilde s_{\rm NH}^2
=&-\tau ^2 \tilde r^2 (1+\tilde t\tilde r)^{-n_T/2}\D
 \tilde t^2
\nonumber \\
&+
\tilde r^{-2} (1+\tilde t\tilde r)^{n_T/2} \left(\D \tilde r^2+\tilde
 r^2\D\Omega_2^2\right)\,.
\label{NHmetric}
\end{align}
As a direct corollary of the scaling limit~(\ref{NHlimit}), 
the near-horizon metric~(\ref{NHmetric}) admits 
a Killing vector
\begin{align}
 \xi ^a=\tilde t\left(\frac{\partial }{\partial \tilde t}\right)^a
-\tilde r\left(\frac{\partial }{\partial \tilde r}\right)^a\,.
\label{generator}
\end{align}
By a straightforward calculation, we find that the 
{\it above Killing field
is hypersurface-orthogonal}
\begin{align}
\xi_{[\mu }\ma D_\nu \xi_{\rho ] }=0\,,
\end{align}
where $\ma D_\mu $ is a derivative operator of the near-horizon metric~(\ref{NHmetric}).
Hence, by transforming to the new coordinates  
\begin{align}
&\tilde T=\pm \ln  |\ti t|+\int^{\tilde R}\frac{4\tilde R^{3+4/n_T}\D
 \tilde R}{n_T(\tilde R^{4/n_T}-1) f(\tilde R)}\,, 
\nonumber \\
&\tilde R=(1+\tilde t \tilde r)^{n_T/4}\,,
\end{align}
where 
\begin{align}
f(\ti R):=\tau^2 \left(\ti R^{4/n_T}-1\right)^2-\ti R^4\,,
\label{f(R)}
\end{align}
we can bring the near-horizon metric~(\ref{NHmetric}) into a manifestly static form,
\begin{align}
 \D \tilde s^2=-\frac{f(\tilde R)}{\tilde R^2}\D \tilde T^2+\frac{16\tau
 ^2 \tilde R^{8/n_T}}{n_T^2f(\tilde R)}\D \tilde R^2+\tilde R^2\D
 \Omega_2^2\,.
\label{NHmetric2}
\end{align}
In this coordinate system, the Killing vector~(\ref{generator})
is expressed as $\xi^\mu =(\partial /\partial \ti T)^\mu $. 
It is obvious that the above metric~(\ref{NHmetric2}) 
describes a static spacetime with Killing horizons at 
$f(\tilde R)=0$ where the Killing vector $\xi^\mu $ becomes null.  
It deserves to note that terms 
in square bracket in Eq.~(\ref{Rconst}) simplify to $f(\ti R)$
in the near horizon limit.
This means that the trapping horizons 
tend to be null as approaching $\ti r\to 0$, in accordance with
the results in the previous subsection.

Since $\ti t$ and $\ti r$ are not gauge-invariant quantities, 
one may be worried about whether the scaling limit~(\ref{NHlimit})
is indeed well-defined.  This prescription of ``zooming-up'' can be
justified as follows. 
In the scaling limit~(\ref{NHlimit}), the dilaton and the U$(1)$
 fields reduce to
\begin{align}
\kappa \Phi =\sqrt{\frac{2n_S}{n_T}}\ln \tilde R\,,
\label{NHPhi}
\end{align}
and
\begin{align}
\kappa \ti F_{\ti T\ti R}^{(T)}=\frac{4\sqrt{2\pi}\tau }{n_T\ti
 R^{1+4/n_T}}\,,
\quad ~~
\kappa \ti F^{(S)}_{\ti T \ti R}=\frac{4\sqrt{2\pi}\tau }{n_T
 R^{1-4/n_T}}\,.
\label{NHF}
\end{align}
We can confirm that the spacetime~(\ref{NHmetric2}) with (\ref{NHPhi}) and
(\ref{NHF}) still satisfies the field equations of Einstein-``Maxwell''-dilaton 
gravity (\ref{Einsteineq})--(\ref{Maxwelleq}). This warrants that the 
near-horizon limit (\ref{NHlimit}) is not the singular 
procedure since it maps the original metric~(\ref{metric}) onto
another one~(\ref{NHmetric}) {\it in the same theory}. 
Note that the case $n_T=4$ does not yield the near-horizon metric, 
since the metric~(\ref{NHmetric}) is identical to the original
one~(\ref{metric0}) and 
simply giving rise to a static chart of the RNdS black hole with $M=Q$.

Let us devote some space to discuss the near-horizon geometry in more
detail. Although the near-horizon metric~(\ref{NHmetric}) is neither 
asymptotically flat 
nor asymptotically AdS, the global causal structures reduce to those of
known solutions. 
Such unusual asymptotic structures of black holes are 
commonly found when the dilaton field is 
present~\cite{Chan:1995fr,Yazadjiev:2005du,soda}. 
The near-horizon metric~(\ref{NHmetric}) admits horizons at
$f(\ti R)=0$, 
which is rewritten into the equation 
\begin{align}
F_\pm (\tilde R)=\tau \,,~~
{\rm with}
~~~
F_\pm (\ti R):=\frac{\pm \ti R^2}{\ti R^{4/n_T}-1}\,.
\label{Fpm}
\end{align}
This equation gives the candidates for horizon radius,
$\ti R_{-}$, $\ti R_{+}$, and 
$\ti R_c$, which were found in 
the limit of $r\rightarrow 0$ of
the trapping horizons in the previous subsection.
The number of roots for Eq.~(\ref{Fpm})
[or $f(\ti R)=0$] can be seen visually as 
the intersection points of $F_\pm (\ti R)$ and 
$\tau ={\rm constant}$  (see Figure~\ref{FRfig}).

Irrespective of the values of $\tau $ and $n_T$, 
behavior of $F_-$ is universal: it starts from zero and 
diverges as $\ti R \to 1$. Then,  $F_-(\ti R)=\tau $
has always a  solution $\ti R=\ti R_{-}$ less than unity.
Whereas,  the behavior of $F_+$ depends on the value of
 $n_T$, i.e.,
$n_T<2$, $n_T=2$ or $n_T>2$. 
As is clear from Figure~\ref{FRfig}, 
for $n_T<2$ [Case I in Table~\ref{Table}] and $n_T=2$
with $\tau>1$ [Case II-(i)], we find two horizons 
($\ti R_{-}$ and $\ti R_{+}$).
When $n_T=2$ and $\tau \leq 1$ [Cases II-(ii) and (iii)],
we have only one root ($\ti R_{-}$).
In the case of $n_T>2$, where the background universe is accelerating,
$F_+$ has a minimum at
\begin{align}
 \ti R= \ti R_{\rm cr}:= \left(\frac{n_T}{n_T-2}\right)^{n_T/4}\,.
\end{align} 
Hence, for
\begin{align}
\tau> \tau_{\rm cr}:=\frac 12 n_T^{n_T/2}(n_T-2)^{1-n_T/2}\,.
\end{align}
the equation (\ref{Fpm}) has three distinct roots, 
$\ti R_{-}<\ti R_{+}\le \ti R_c$ [Case III-(i)]. 
When $\tau=\tau_{\rm cr}$, two horizons $\ti R_{+}$ and $\ti R_c$ become
degenerate [Case III-(ii)].  
If $\tau< \tau_{\rm cr}$, we have only one root [Case II-(iii)].
These results are shown in Table \ref{Table}. 
\begin{widetext}
\begin{center}
\begin{figure}[h]
\includegraphics[width=15cm]{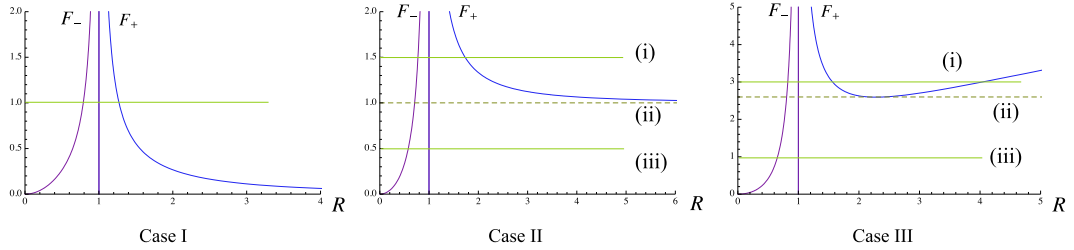}
\caption{Plots of $F_+$ and $F_-$  for 
Case I (left),  Case II (middle) and Case III (right)
(see Table \ref{Table} for classification). 
The equality  $f(\ti R)=0$ then 
possesses two distinct solutions for Case I
and Case II-(i), a single solution 
for Cases II-(ii), II-(iii) and III-(iii),  
three distinct solutions for Case III-(i) and two solutions
with larger one  being degenerate for Case III-(ii).}
\label{FRfig}
\end{figure}
\end{center}
\end{widetext}

The spacetime structure of the near-horizon metric for each case
I--III is obtained by simple evaluation of radial null geodesics
 (see~e.g.,~\cite{TM}). 
We will not devote much time to discussing this issue since our primary 
interest in this subsection is exclusively to the horizon structure. 
We obtain 5 types of conformal diagrams,  
as shown in Figure~\ref{PDstatic} (A)--(E).  
The figure (A) corresponds to
Cases I and II-(i), (B) to Case II-(ii), (C) to Case 
II-(iii) and III-(iii), (D) to Case III-(i) and (E) to Case III-(ii).   
These classifications are also included in Table~\ref{Table}. 
Despite the nontrivial asymptotic structures~(\ref{NHmetric2}), these 
conformal diagrams are identical to the familiar solutions. 
The figure (A) is the same as the RN-AdS, (B) as the vanishing mass RN
spacetime with a hyperbolic horizon, (C) as the negative mass SdS, 
(D) as the nonextremal RNdS and (E) as the degenerate RNdS, respectively.
The surfaces $\ti R_{+}$ and $\ti R_{-}$ are the black hole event
horizon and the white hole horizon for the near-horizon geometry 
and also for the original metric.   It should be 
emphasized, however, that the largest horizon $\ti R_c$ does 
{\it not} approximate our original
metric, since 
the surface $\ti R_c$ corresponds to the limit $\ti t\to 0$
and $\ti r\to \infty $  with $\ti t \ti r $ being finite in the 
near-horizon geometry
(the green line in Figure~\ref{PDstatic}). It incidentally arises 
due to the symmetry of the near-horizon metric not shared by 
the original metric. 
Consequently, 
the null surface $\ti R_c$ is irrelevant to the 
cosmological horizon in the original metric~(\ref{metric0}), although
it {\it is} a cosmological horizon for the near-horizon metric. 
It is worth commenting also that the case $n_T=n_S=2$ with $\tau=1$ is 
special, 
for which there exists ``internal null infinity'' $\ma I^+_{\rm in}$ 
specified by $\ti t \to \infty $ and $\ti r\to 0$ with 
$\ti R=(\ti t\ti r+1)^{1/2}$
being infinite. This surface has an outgoing null structure. 
Only the ingoing null geodesics can approach $\ma I^+_{\rm in}$, 
hence it should be distinguished by $\ma I^+_{\rm out}$ 
($\ti t\to \infty $ and $\ti r\to \infty $).   
The Carter-Penrose diagram in this case is given in 
Figure~\ref{PDstatic}(B).

To conclude, the neighborhoods of the null surfaces $\ti R_{+}$ 
and $\ti R_{-}$ in the original metric~(\ref{metric0})
are locally isometric to the corresponding null surfaces--these
are indeed the Killing horizons--in the near-horizon metric 
(\ref{NHmetric}) or (\ref{NHmetric2}). 
In Figure~\ref{PDstatic}, only the shaded regions in the vicinity 
of  the horizon approximate our original metric~(\ref{metric0}). 
The surface $\ti R_{+}$, if exists, deserves a black-hole event horizon
for the near-horizon metric and it turns out later, this surface
corresponds to a black hole horizon for the original metric as well.

Note also that the 
vector field $\xi^\mu $~(\ref{generator}) solves the Killing equation in the 
original spacetime only at the horizon. Since the outside region is
highly dynamical, 
it comes as a surprise for us that the unexpected Killing symmetry
arises. This is achieved due to the fact that the electromagnetic
repulsive force and the cosmic expansion are compensated 
by the gravitational and dilatonic attraction. 
These Killing horizons $\ti R_\pm$ are in general nondegenerate 
in the sense that  
the surface gravities $\ti \kappa_\pm$ are nonvanishing. 
From the formula 
\begin{align}
2\ti \kappa^2_\pm =\mp (\ti\nabla_\mu \xi_\nu )(\ti \nabla^\mu \xi^\nu )\,,
\end{align}
one obtains the surface gravities associated with $\xi^\mu $ as 
\begin{align}
\ti \kappa_ \pm =1\mp  \frac{n_T \ti R^{2-4/n_T}_\pm}{2\tau }\,.
\label{kappapm}
\end{align}
These are nonvanishing unless $\tau=\tau_{\rm cr}$.
The surface gravities are constant over the horizon, as expected.

\subsubsection{Null geodesics}

The argument given in  previous subsection 
strongly implies that $\ti R_{+}$ and $\ti R_{-}$
are the future and past event horizons. 
In order to conclude this in a more rigorous manner, 
we need to discuss geodesic motions. 
Since the present spacetime admits a spherical symmetry, 
it is sufficient to focus on the radial null geodesics
to determine the causal structure.

\begin{widetext}
\begin{center}
\begin{figure}[t]
\includegraphics[width=12cm]{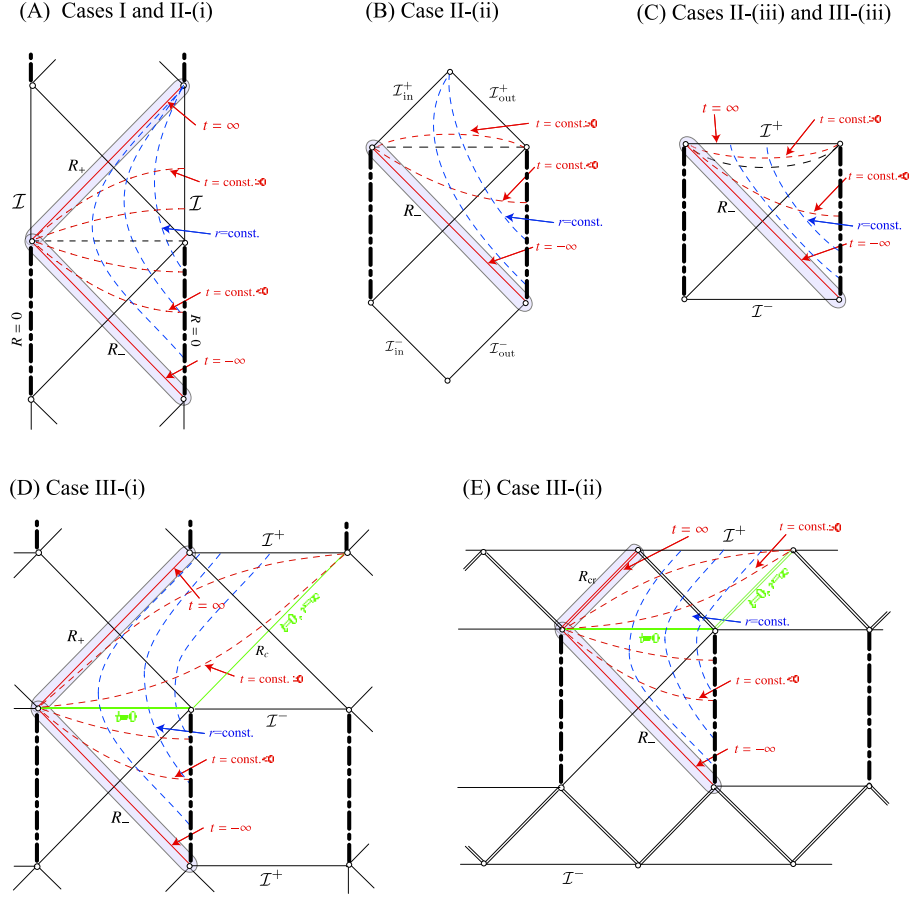}
\caption{Conformal diagrams of the maximally extended 
near-horizon geometries.  The global structures are the same as 
those for 
(A) the nonextremal RN-AdS, (B) the zero-mass RN spacetime 
with a ``hyperbolic horizon,'' (C)  the negative mass SdS
(or the overcharged RNdS), (D) the nonextremal RNdS and (E) the extremal RNdS. 
The thick dotted-dashed lines denote the central singularity 
at $\ti t=-1/r$. 
The shaded regions approximate our original solution~(\ref{metric0}).}
\label{PDstatic}
\end{figure}
\end{center}
\end{widetext}

Using the relation 
for the radial null vectors,  
\begin{align}
\tau \dot{\tilde t}= \pm \sqrt{
H_T^{n_T}
H_S^{n_S}
}\dot{\tilde r}\,,
\label{nulleq0}
\end{align}
the radial null geodesic equations 
simplify to 
\begin{eqnarray}
\ddot{\tilde{t}}\pm \frac{\tau H_{ST}}{2\ti r^2(H_T^{n_T+2}
H_S^{n_S+2})^{1/2}} \dot{\tilde{t}}^2&=&0\,, 
\label{nulleq2} \\
\ddot{\tilde r}\pm \frac{n_T (H_T^{n_T}H_S^{n_S})^{1/2}}
{2\tau H_T}\dot{\tilde r}^2&=&0\,,
\label{nulleq3}
\end{eqnarray}
Here and hereafter, the plus (minus) sign refers to as the
outgoing (ingoing) null geodesics, and the dot
means the differentiation with respect to an affine parameter
$\lambda$. 

Following~\cite{MN,Brill} the asymptotic geodesic behavior 
around the surface $\ti R_+$ (corresponding to 
$\ti t \to \infty $ and $\ti r\to 0$) can be evaluated as 
\begin{align}
\ti r =c_1^{(+)} (\lambda-\lambda_+)^{1/\ti \kappa_+}\,, \qquad
\ti t =c_2^{(+)} (\lambda-\lambda _+)^{-1/\ti \kappa _+}\,,
\end{align} 
where $c_{1,2}^{(+)}$ is constant, 
$\lambda_+$ corresponds to the arrival time at $\ti R_+$ and 
$\ti \kappa_+$ was defined by Eq.~(\ref{kappapm}). 
This equation not only means that the horizon can be reached within a 
finite affine time but also that $\ti r$ and $1/\ti t$ are not the smooth
functions of $\lambda $ at $\lambda _+$
(unless $\ti \kappa_+^{-1}$ happens to be integral).

Let us move on to the numerical analysis. 
Exploiting the
affine transformation, if necessary, 
we can impose without loss of any generality 
$\lambda =0$ at the starting point of the geodesics 
and $\dot{\tilde t}(0)=\pm 1$.   
Then the geodesics are distinguished by their initial
point $[\ti t(0), \ti r(0)]$.

Our primary concern here is Case III, 
where the background universe is accelerating. 
We will give a detailed description exclusively of  
$n_T=3$ with $\tau=3$ [Case III-(i)]. 
We first develop geodesics for the outside region $\ti r>0$
by setting the initial radial coordinate to be  
$\ti r (0)=1$.  It then follows that the $\ti r=1$ surface is 
divided by three trapping horizons into four regions (see Figure~\ref{fig:Rr}). 
Take the representative spacetime points  
$p_I~(I=1,...4)$ specified by $(\ti t_I(0), \ti r(0)=1)$
with $t_1(0)>t_2(0)>t_3(0)>t_4(0)$ in such a way that 
$p_I$ is contained in each partitioned (un)trapped region.  
We shall refer to the geodesics starting from the initial point $p_I$ 
as ``Class-$I$.''   
For $\ti r<0$, we take $\ti r(0)=-1/10$ and repeat the identical procedure 
with redefining ``Class-$I$'' as 
$t_4(0)<t_3(0)<t_2(0)<t_1(0)$.  
Figure~\ref{fig:geo} presents our numerical result 
in the accelerating universe ($n_T=\tau=3$) and the deduced Carter-Penrose diagrams
are shown in Figure~\ref{PD}(d).

\begin{widetext}
\begin{center}
\begin{figure}[h]
\includegraphics[width=15cm]{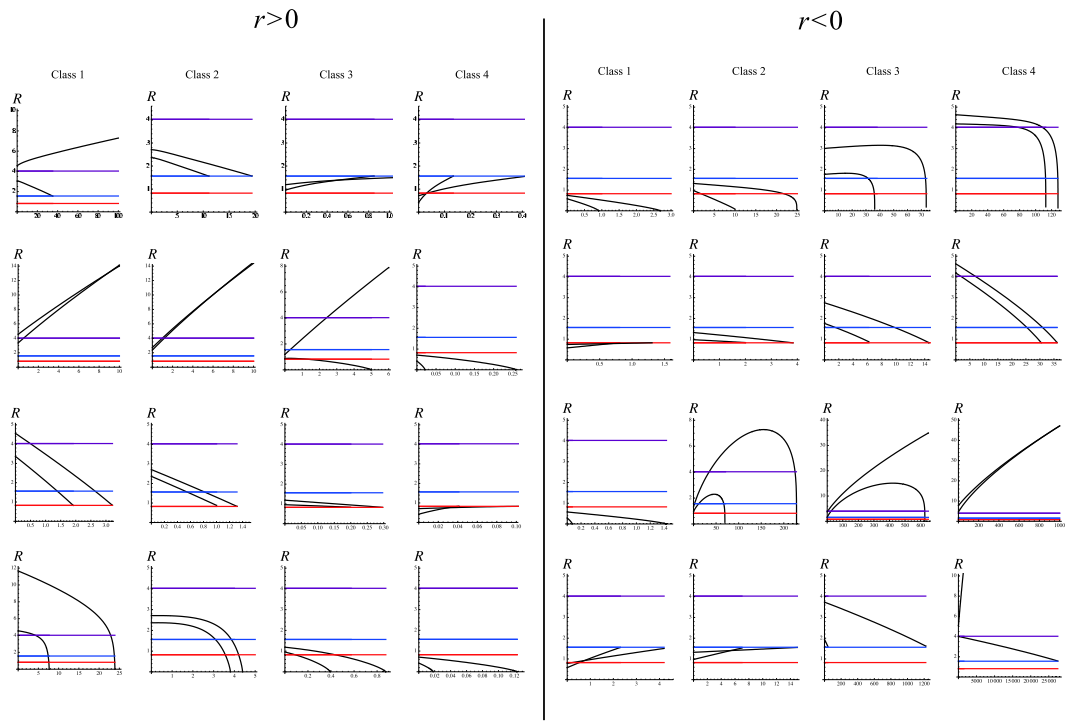}
\caption{Behaviors of radial null geodesics 
for the accelerating universe ($n_T=3$ with $\tau=3$) in the outside
 region (left) and the inside region (right). The figures show  
the circumference radius versus the affine parameter. The red, blue and 
purple lines  stand for $\ti R_{-}\sim 0.82$, $\ti R_{+}\sim 1.56$ and $\ti R_c\sim 4.03$,
 respectively. The diagrams in the top, upper  middle, lower middle, bottom
rows correspond to future-directed ingoing null geodesics, 
future-directed outgoing null geodesics, past-directed ingoing null
 geodesics and past-directed outgoing null geodesics.
We have shown the null geodesics for the initial values
 $\ti r(0)=1 $ (left) and $\ti r(0)=-1/10$ (right). }
\label{fig:geo}
\end{figure}
\end{center}
\vspace{-1cm}
\end{widetext}

Let us begin by the analysis for $\ti r>0$. 
Among the Class-1 geodesics of the future-directed ingoing null,  
geodesics with a sufficiently large $\ti t_1$ tend to diverge 
as the affine time evolves, while 
geodesics with $\ti t_1(0)$ being close to the trapping horizon  
converge to $\ti R_+$ with an infinite redshift.   
This implies the existence of a cosmological horizon   
[the $45^\circ$ line denoted by CH in Figure~\ref{PD}(d)], which is a 
characteristic feature in the accelerating universe as seen
for the $p>1$ case in Figure~\ref{fig:FRW}.  
Other classes of geodesics unavoidably approach $\ti R_+$
within a finite affine parameter.

Behaviors of future-directed outgoing null geodesics of Class-3
imply the existence a critical null curve 
[the black dotted $45^\circ $ line in Figure~\ref{PD}(d)]
above which the geodesics 
can reach infinity and below which the geodesics 
inevitably plunge into the singularity 
$\ti t=\ti t_s(\ti r)$. 

Motions of the past-directed ingoing null geodesics
are universal: they necessarily get to the null surface $\ti R_{-}$
with undergoing an infinite blueshift.

The past-directed outgoing null geodesics
also exhibit a universal behavior: they inevitably 
reach the timelike singularity 
$\ti t_s(\ti r)$ [the gray dashed-dotted line $\ti R=0$ in Figure~\ref{PD}(d)]
within a finite affine parameter.

\begin{widetext}
\begin{center}
\begin{figure}[t]
\includegraphics[width=12cm]{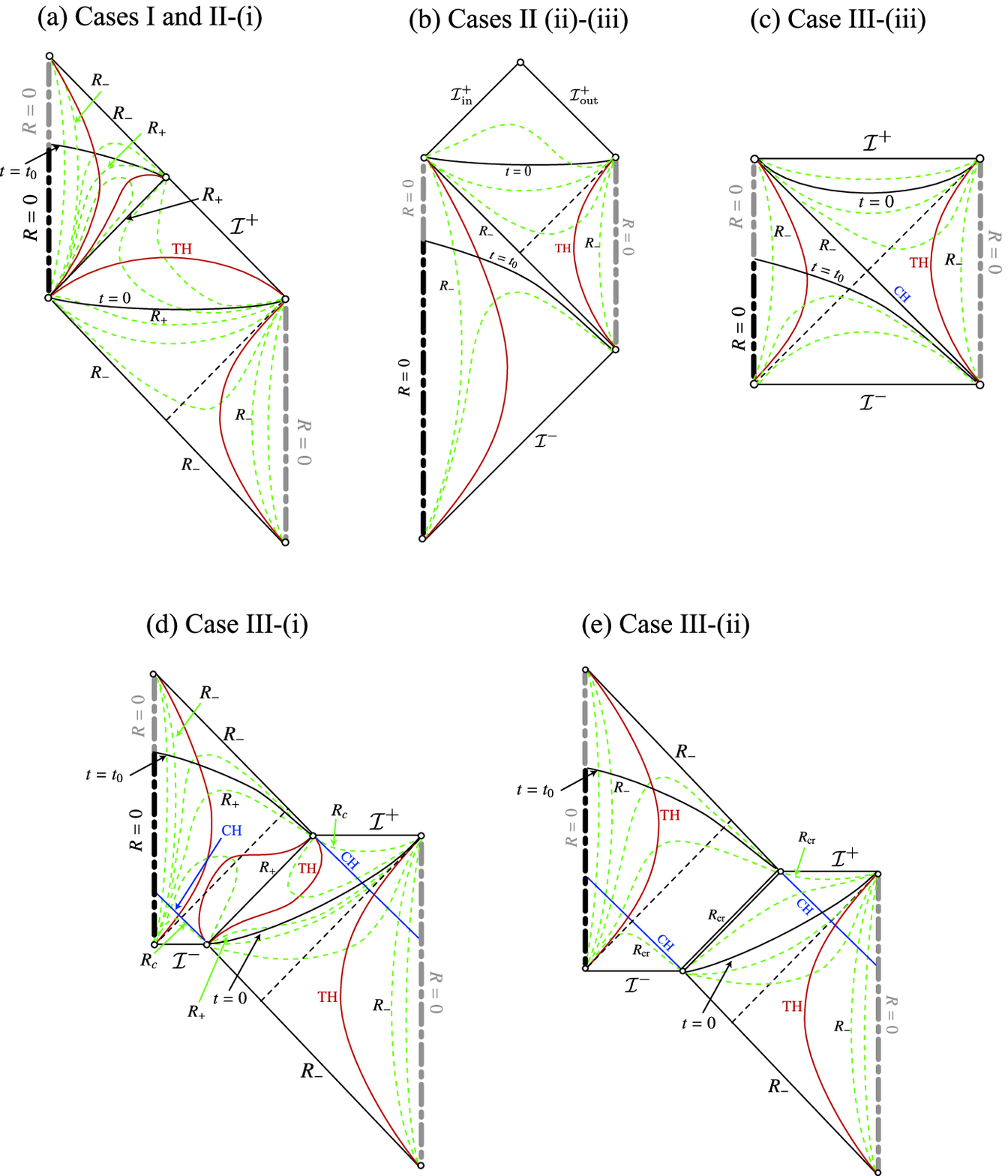}
\caption{Conformal diagrams of the present spacetime~(\ref{metric})
for Cases I--III.  The green dotted curves stand for the
$R={\rm constant}$ surfaces.
The red thick curves represent the trapping
horizons (TH) and blue lines denote the cosmological horizons (CH). 
The gray and black dotted-dashed curves correspond to 
the singularities of $r=-Q$ and $t=t_s(r)$, respectively.   
Figure (a), where the universe 
is decelerating, has been argued in the previous
paper~\cite{MN}. For Figure (b), the universe undergoes a
marginal acceleration $a \propto \bar t$.
It admits an internal null infinity  
$\ma I^+_{\rm in }$ at which $r=0$ and $t \to \infty $ with 
$R\to\infty $. 
Figures (c)--(e) correspond to the accelerating universe. 
Figure (d) shows that there appear three horizons:
the black hole horizon $R_{+}$, 
the ``white hole horizon'' $R_{-}$ and the 
cosmological horizon denoted by CH (not $R_c$).  In Figure (e), 
the horizon is degenerate, while in Figure (c) the spacetime is 
overcharged to give a globally naked singularity.  
Figures~(a), (d) and (e) are extensible across the surface
 $R_{-}$ with a contracting universe 
in a continuous but nonanalytic manner.  
}
\label{PD}
\end{figure}
\end{center}
\end{widetext}

Assembling these results, we are able to draw a 
conformal diagram of Figure~\ref{PD}(d).\footnote{
We traced several geodesics starting from different initial points
and  checked consistency.
}
The future infinity consists of a spacelike surface, 
consistent with the behavior of future-directed ingoing null geodesics
of Class-1, and with the FLRW limit $\ti r\to \infty $ 
as described in Figure~\ref{fig:FRW}(V). 
The timelike singularity $\ti t _s(\ti r)=-1/\ti r$ 
exists up to $\ti t=0$, which is joined with $\ma I^+$.  
The cosmological horizon develops, 
whose orbit $\ti t=\ti t_{\rm CH}(\ti r)$ can be traced numerically. 
The green-dotted lines denote the $\ti R={\rm constant}$
surfaces, which change signature across the trapping horizons. 
The $\ti R={\rm constant}$ surfaces can be written in comparison with
Figure~\ref{fig:Rr}, and the trapping horizons can be drawn from 
Eq.~(\ref{metric_TH2})
and (\ref{metric_TH1}). 
The Carter-Penrose diagram explicitly shows that the null surface 
$\ti R_{+}$
deserves to be a black hole horizon. 
The null surface $\ti R_{-}$, on the other hand, may not suitable 
to be called a white hole horizon in a rigorous sense since 
the spacetime admits no past infinity in the outside region.
Nevertheless, we shall stick to call it a ``white hole horizon,''  
from the spirit of a one way membrane as a ``region of no-entrance.''

The global structure of the solution
in the inside region ($\ti r<0$) can be deduced analogously. 
It is notable that the inside region admits a 
past null infinity $\ma I^-$.
At first glance, it seems peculiar because 
the background universe is expanding 
(at $\ma I^-$ the circumference radius $\ti R$ is already infinite there
and the universe is not able to expand further!).  
We recognize that the spacetime indeed approaches to 
the expanding FLRW universe only in the limit of $r \to +\infty $ with 
$t/t_0>0$ [see Eq.~(\ref{tbar})],  while this is not true  
in the limit $t/t_0\to -\infty $ with $-1<\ti r<0$, 
in which case the background geometry looks like a
contracting universe.

Following the above treatment, 
one can draw the Carter-Penrose diagrams for other cases.
We do not display those behaviors of null geodesics corresponding to 
Figure~\ref{fig:geo}
in order to retain the compactness of the present paper, but 
they are easy to obtain. (Without geodesic analysis 
one can intuitively deduce the causal structures  simply 
by inspecting the $\ti R-\ti r$ diagram in Figure~\ref{fig:Rr}). 
We have arrived at 5-types of global conformal diagrams shown in 
Figure~\ref{PD} (a)--(e). Let us summarize the results.

Figure~\ref{PD}(a) corresponding to Cases I and II-(i) 
was given in our previous paper~\cite{MN}. 
The spacetime has a regular, nondegenerate event horizon $\ti R_{+}$,  and
approaches to the decelerating FLRW universe. 
In the picture shown in Fig~\ref{PD}(a), we 
sketch the 
trapping horizons for the $n_T=\tau=1$ case, where the spacelike 
trapping horizon $\ti R_1$
develops in the outside domain. Its signature depends sensitively 
on $n_T$ and $\tau $. For example, it becomes asymptotically null
in the radiation-dominated case ($n_T=4/3$). 
Though the signature of trapping horizons is modified depending on the
parameters, the causal structure remains unaltered.

In Figure~\ref{PD}(e) [Case III-(ii)], 
the spacetime has a degenerate event horizon at 
$\ti R_{+}=\ti R_c$ and asymptotically tends to
the accelerating FLRW universe. 
This is a special situation of Case III-(i).

On the other hand, the spacetime fails to have a black hole horizon in
Figures~\ref{PD}(b) [Cases II-(ii) and II-(iii)] and \ref{PD}(c) 
[Case III-(iii)] (see Table~\ref{Table}). 
These cases correspond to  spacetimes with naked singularities.  
In Figure~\ref{PD}(b) where $n_T=n_S=2$, one may recognize that there exists
the ``internal'' null infinity 
$\ma I^+_{\rm in}$, where $\ti r\to 0 $ and $\ti t \to \infty $
with $\ti R\sim (\ti t \ti r)^{1/2} \to \infty $. 
We can show the existence of $\ma I^+_{\rm in}$ as follows.
Setting $n_T=n_S$ and assuming $\ti r\sim 0$ with $\ti t \to \infty $, 
one can find the asymptotic solutions of the
radial ingoing null geodesics~(\ref{nulleq2}) and (\ref{nulleq3}) around
 $\lambda \to \infty $ as
\begin{align}
r=\lambda ^{-\tau/(1-\tau )}\,, ~~~ &{\rm with}~~~ 
t=C_1\lambda ^{1/(1-\tau )}\,,
\end{align}
for $\tau<1$, and 
\begin{align}
r=\frac{W(C_2\lambda )}{\lambda }\,, 
~~~ &{\rm with}~~~ t=\lambda \,,
\end{align}
for $\tau =1$, 
where $C_1$ and $C_2$ are positive constants and $W$ is 
Lambert's $W$-function satisfying 
$W (x)e^{W(x)}=x$. These asymptotic geodesics both satisfy 
$\ti t \ti r\to \infty $ as $\lambda \to \infty $. 
Whereas, the outgoing radial null geodesics do not decrease $\ti r$. 
This  proves the existence of $\ma I^+_{\rm in}$, at which only the
ingoing null geodesics can arrive.

We have revealed that our solution~(\ref{metric0}) 
specified by three continuous parameters ($Q, n_T, \tau $) describes a 
charged black hole in the FLRW universe in some parameter range.
It is therefore worthwhile to discuss the physical meaning of these
parameters. It is clear that the constant $Q$ is 
a U$(1)$ charge, whilst 
$n_T$ is the steepness of the dilaton potential 
(if $n_T$ happens to be natural number, it counts the number of
 time-dependent branes).

\begin{figure}[t]
\begin{center}
\includegraphics[width=6cm]{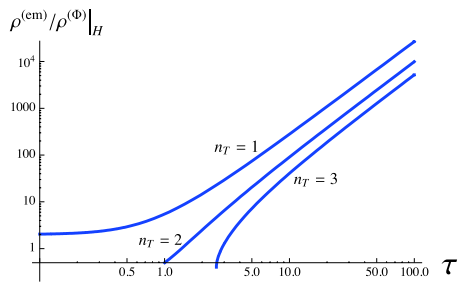}
\caption{
Relation between $\tau$ and 
the ratio of two energy densities at the horizon in the double 
logarithmic plot. 
This illustrates the almost linear relationship between 
$\tau^2 $ and $\rho^{\rm (em)}/\rho^{(\Phi)}|_H$. 
}
\label{rhotau}
\end{center}
\end{figure}

If $n_T=4$, the parameter $t_0$ has a clear meaning
as a curvature radius of the de Sitter spacetime.
However,  the meaning of $\tau $ (or $t_0$) is less obvious for $n_T \ne 4$.
In order to clarify the meaning,
 we shall consider the energy densities of two fields. 
The values of $\rho^{(\Phi)}$ and $\rho^{\rm (em)}$ on the 
event horizon $\ti R_{+}$
are given by
\begin{eqnarray}
\kappa^2 \rho^{(\Phi)}|_H&=&{n_T(n_T+2)\over 8 t_0^2}
\ti R_{+}^{-{2n_S/ n_T}}
\,,
\\
\kappa^2 \rho^{\rm (em)}|_H&=&
{n_S+n_T\ti R_{+}^{-{8/ n_T}}
\over 4Q^2 \ti R_{+}^{2}}
\,,
\end{eqnarray}
which leads  to
\begin{eqnarray}
{\rho^{\rm (em)}\over \rho^{(\Phi)}}\Big{|}_H
={2\tau^2\over n_T(n_T+2)}
\ti R_{+}^{-4}
\left(
n_T+n_S\ti R_{+}^{{8/ n_T}}
\right)
\label{tau_meaning}
\,.
\end{eqnarray}
The horizon radius $\ti R_{+}$
is given in terms of  $\tau$ via Eq.~(\ref{Fpm}).
As a result, we find that  
the ratio of two energy densities at the horizon is related to
$\tau$,
although the relation is complicated.
In the limit of $\tau\rightarrow \infty$,
however, we find a simple relationship
\begin{eqnarray}
{\rho^{\rm (em)}\over \rho^{(\Phi)}}\Big{|}_H
\rightarrow 
{8\over n_T(n_T+2)}\,
\tau^2
\,.
\end{eqnarray}
Figure~\ref{rhotau} depicts the dependence of
${\rho^{\rm (em)}/\rho^{(\Phi)}}$ against $\tau $.
It follows that $\tau^2$ has a one to one correspondence to  
the ratio of two energy densities evaluated at the horizon $\ti R_+$.

We can also give an afterthought implication to $\tau$ as follows.
From Eq.~(\ref{Fpm}), one obtains a relationship
\begin{align}
\tau =
\frac{\ti R_{+}^2+\ti R_{-}^2}{\ti R_{+}^{4/n_T}-\ti R_{-}^{4/n_T}}\,.
\end{align} 
This equation implies that the parameter $\tau$
measures the ``nonextremality'' of the Killing horizon. 
An extremal limit is taken $\tau \to \infty $ 
where $\ti R_{+}=\ti R_{-}$ and 
$\tau \to \tau_{\rm cr}$ where $\ti R_{+}=\ti R_c$. 
Inferring from the supersymmetric case,
the nonextremality indicates that the mass is strictly larger than 
the (central) charge. Hence, we deduce that some combination of 
the above parameters will give a well-defined mass of a 
black hole larger than the charge $Q$. 
Considering that the black hole horizon does not occur for small $\tau $, 
this point of view is supported.
If a well-behaved mass definition is found, 
we are able to discuss the thermodynamics of a black hole.  
This expected possibility is now under investigation.

\section{Concluding Remarks}
\label{sec:summary} 

We have presented a family of solutions in a $D$-dimensional 
Einstein-``Maxwell''-dilaton system with a Liouville potential. 
In the single mass case, the solution  
interlies the extremal black hole and the FLRW universe with 
a power-law expansion. This solution reduces in the special cases
to the extremal RN black hole ($n_T=0$), the nonextremal RNdS black hole
($n_S=0$) and the solution of~\cite{MOU} derived from 
the intersecting M-branes ($D=4$ and $n_T=1$).   
The present spacetime is characterized by
three parameters: the steepness of the dilaton potential $n_T$, the U$(1)$
charge $Q$ and the ``nonextremality''
(or the ratio of two energy densities at the horizon) $\tau $.  
The parameter $n_T$ controls the expanding power $p$ of the background FLRW universe. 
For any choice of parameters, the system is shown to obey the weak energy condition.

The primary aim of this paper is to obtain global structures of the
solution.  The spacetime can be grouped into the 
nine cases 
[Cases I-(i)--III-(iii)]
 summarized in Table~\ref{Table}, according to
the parameter values.   
Our consequence is visually captured in Figure~\ref{PD}, where
the 5-types of global structure are obtained. 
In the case where the background universe is accelerating ($n_T>2$)
with large nonextremality parameter ($\tau\ge \tau_{\rm cr}$), 
the spacetime indeed describes a charged black hole in
the FLRW universe undergoing an accelerated expansion [Case III-(i)]. 
We have also clarified the global structures for the 
marginally accelerating and decelerating cases. 
This has been an open issue in~\cite{GMII}.

The present solution displays various interesting features. 
At first glance, the solution appears to have a degenerate horizon. 
We have shown that this is not the case. Taking the near-horizon
limit~(\ref{NHlimit}), the  horizon is not degenerate except
 for $\tau=\infty $
and $\tau=\tau_{\rm cr}$.   
Amazingly, the event horizon of a black hole constitutes a 
Killing horizon. This means  that the 
matter fields fail to accrete into the black hole, whereby
the area of the black hole keeps constant.  Hence, 
the present spacetime is not suitable for describing a growing black hole in 
FLRW universe. Instead, 
the solution preserves the analogue of the equilibrium state 
despite the time-dependence of the metric. 
The situation is closely analogous to the supersymmetric state.

The black hole thermodynamics in the expanding universe is an interesting 
future work to be discussed. 
Since the event horizon is generated by a (partial)
Killing field, the black hole has a nonvanishing surface gravity. 
However, it is far from obvious to which observers the physical 
temperature is assigned, since the surface gravity
is sensitive to the normalization of a Killing field. 
This may be rephrased as what is the mass of the black hole. 
We hope to visit this issue in a separated paper.

Another interesting issue to be explored in the time-dependent black hole
spacetime is a black hole merger.
Adding multiple point sources and reversing time backwards, 
the present metric turns out to give the analytic description of 
black hole coalescence, similar to the Kastor-Traschen 
solution~\cite{KT,HH}. 
In the single-mass Kastor-Traschen spacetime (RNdS), 
an analogue of ``thermal equilibrium'' is realized due to the 
fact that the temperature of the event horizon and that of the cosmological
horizon become the same. If this viewpoint continues to be valid in the
present spacetime, what plays the role of a box containing the black
hole which is thermal equilibrium with a bath of radiation?  
These are interesting questions to be explored.

\acknowledgments

M.N would like to thank Nobuyoshi Ohta for discussions. 
This work was partially supported 
by the Grant-in-Aid for Scientific Research
Fund of the JSPS (No.19540308) and for the
Japan-U.K. Joint Research Project,
and by the Waseda University Grants for Special Research Projects.



\begin{thebibliography}{99}

\bibitem{PBH1}
  Y.B. Zeldovich and I.D. Novikov, Sov. Astr. AJ  {\bf 10}, 602 (1967).

\bibitem{PBH2}
  S.~W.~Hawking,
  Mon.\ Not.\ Roy.\ Astron.\ Soc.\  {\bf 152}, 75 (1971): 
  B.~J.~Carr and S.~W.~Hawking,
  Mon.\ Not.\ Roy.\ Astron.\ Soc.\  {\bf 168}, 399 (1974).

\bibitem{PBHs}
  B.~J.~Carr,
  arXiv:astro-ph/0511743; 
  B.~Carr, K.~Kohri, Y.~Sendouda and J.~Yokoyama,
  arXiv:0912.5297 [astro-ph.CO].

\bibitem{Hawking1974}
  S.~W.~Hawking,
  Commun.\ Math.\ Phys.\  {\bf 43}, 199 (1975)
  [Erratum-ibid.\  {\bf 46}, 206 (1976)].

\bibitem{Carter}
B. Carter, {\it Black Hole Equilibrium States} in {\it Black Holes}, edited
by C. DeWitt and J. DeWitt 
(Gordon and Breach, New York, 1973).

\bibitem{uniqueness}
  M.~Heusler,
  Living Rev.\ Rel.\  {\bf 1}, 6 (1998).


\bibitem{Einstein:1945id}
  A.~Einstein and E.~G.~Straus,
  Rev.\ Mod.\ Phys.\  {\bf 17}, 120 (1945).


\bibitem{SultanaDyer}
J. Sultana and C. C. Dyer, 
Gen.\ Rel.\ Grav. {\bf 37}, 1349 (2005).


\bibitem{HMC}
  T.~Harada, H.~Maeda and B.~J.~Carr,
  Phys.\ Rev.\  D {\bf 74}, 024024 (2006)
  [arXiv:astro-ph/0604225].




\bibitem{MOU}
  K.~Maeda, N.~Ohta and K.~Uzawa,
  JHEP {\bf 0906}, 051 (2009)
  [arXiv:0903.5483 [hep-th]].


\bibitem{MN}
K. Maeda and M. Nozawa, 
Phys.\ Rev.\ D {\bf 81}, 044017 (2010)
[arXiv:0912.2811 [hep-th]].



\bibitem{GMII}
G. W~Gibbons and K.~Maeda,
arXiv:0912.2809 [gr-qc], 
accepted for publication in Phys.\ Rev.\ Lett.

\bibitem{Mcvittie1933}
G. C.~McVittie,
Mon. Not. R. Astron. Soc.{\bf 93}, 325 (1933).


\bibitem{Nolan} 
B. C. Nolan,
Phys. Rev. D{\bf 58}, 064006 (1998);
Class. Quantum Grav. {\bf 16} 1227 (1999);
Class. Quantum Grav. {\bf 16} 3183 (1999).


\bibitem{Carrera:2009ve}
  M.~Carrera and D.~Giulini,
  arXiv:0908.3101 [gr-qc].


\bibitem{Kaloper:2010ec}
  N.~Kaloper, M.~Kleban and D.~Martin,
  arXiv:1003.4777 [hep-th].

\bibitem{noncritical_string_theory}
  E.~Witten,
  Phys.\ Rev.\  D {\bf 44}, 314 (1991).

\bibitem{massiveIIA}
  L.~J.~Romans,
  Phys.\ Lett.\  B {\bf 169}, 374 (1986).


\bibitem{Kaluza-Klein_compactification}
  I.~V.~Lavrinenko, H.~Lu and C.~N.~Pope,
  Nucl.\ Phys.\  B {\bf 492}, 278 (1997)
  [arXiv:hep-th/9611134];
  C.~M.~Chen, P.~M.~Ho, I.~P.~Neupane, N.~Ohta and J.~E.~Wang,
  JHEP {\bf 0310}, 058 (2003)
  [arXiv:hep-th/0306291].

\bibitem{VEV_form}
  J.~E.~Lidsey, D.~Wands and E.~J.~Copeland,
  Phys.\ Rept.\  {\bf 337}, 343 (2000)
  [arXiv:hep-th/9909061].





\bibitem{Lucchin:1984yf}
  F.~Lucchin and S.~Matarrese,
  Phys.\ Rev.\  D {\bf 32}, 1316 (1985).

\bibitem{Penrose:1969pc}
  R.~Penrose,
  Riv.\ Nuovo Cim.\  {\bf 1}, 252 (1969)
  [Gen.\ Rel.\ Grav.\  {\bf 34}, 1141 (2002)].

\bibitem{HH}
  J.~H.~Horne and G.~T.~Horowitz,
  Phys.\ Rev.\  D {\bf 48} (1993) 5457
  [arXiv:hep-th/9307177].

\bibitem{MS}
  T.~Maki and K.~Shiraishi,
  Class.\ Quant.\ Grav.\  {\bf 10}, 2171 (1993).



\bibitem{Hartle:1972ya}
  J.~B.~Hartle and S.~W.~Hawking,
  Commun.\ Math.\ Phys.\  {\bf 26}, 87 (1972).





\bibitem{KT}
D. Kastor and J. Traschen, 
Phys. Rev. D {\bf 47}, 5370  (1993)
[arXiv: hep-th/9212035].

\bibitem{London}
  L.~A.~J.~London,
  Nucl.\ Phys.\  B {\bf 434} (1995) 709.

\bibitem{Nariai}
H. Nariai~, 
Sci.\ Rept.\ Tohoku \ Univ.\ (Ser.A) {\bf 34}, 160 (1950); 
Sci.\ Rept.\ Tohoku \ Univ.\ (Ser.A) {\bf 35}, 62 (1951). 

\bibitem{BR}
  B.~Bertotti,
  Phys.\ Rev.\  {\bf 116} (1959) 1331;
  I.~Robinson,
  Bull.\ Acad.\ Pol.\ Sci.\ Ser.\ Sci.\ Math.\ Astron.\ Phys.\  {\bf 7} (1959) 351.


\bibitem{Kunduri:2007vf}
  H.~K.~Kunduri, J.~Lucietti and H.~S.~Reall,
  Class.\ Quant.\ Grav.\  {\bf 24}, 4169 (2007)
  [arXiv:0705.4214 [hep-th]];
\bibitem{Astefanesei:2007bf}
  D.~Astefanesei and H.~Yavartanoo,
  Nucl.\ Phys.\  B {\bf 794}, 13 (2008)
  [arXiv:0706.1847 [hep-th]].


\bibitem{KT2}
  D.~Kastor and J.~H.~Traschen,
  Class.\ Quant.\ Grav.\  {\bf 13}, 2753 (1996)
  [arXiv:gr-qc/9311025].



\bibitem{Behrndt:2003cx}
  K.~Behrndt and M.~Cvetic,
  Class.\ Quant.\ Grav.\  {\bf 20}, 4177 (2003)
  [arXiv:hep-th/0303266].

\bibitem{Freedman:2003ax}
  D.~Z.~Freedman, C.~Nunez, M.~Schnabl and K.~Skenderis,
  Phys.\ Rev.\  D {\bf 69}, 104027 (2004)
  [arXiv:hep-th/0312055].

\bibitem{Grover:2008jr}
  J.~Grover, J.~B.~Gutowski, C.~A.~R.~Herdeiro and W.~Sabra,
  Nucl.\ Phys.\  B {\bf 809}, 406 (2009)
  [arXiv:0806.2626 [hep-th]].



\bibitem{wald2}
R.~M.~Wald, {\it Quantum Field Theory in Curved Spacetime
and Black Hole Thermodynamics},
(University of Chicago Press, 1994).

\bibitem{HE} 
S. W.~Hawking and G. F. R.~Ellis,
{\it The large scale structure of space-time}
(Cambridge: Cambridge University Press, 1973).



\bibitem{NEC}
  R.~V.~Buniy and S.~D.~H.~Hsu,
  Phys.\ Lett.\  B {\bf 632}, 543 (2006)
  [arXiv:hep-th/0502203]; 
  S.~Dubovsky, T.~Gregoire, A.~Nicolis and R.~Rattazzi,
  JHEP {\bf 0603}, 025 (2006)
  [arXiv:hep-th/0512260].



\bibitem{MS1964}
 C.~W.~Misner and D.~H.~Sharp,
  Phys.\ Rev.\  {\bf 136}, B571 (1964).

\bibitem{Hayward1994}
  S.~A.~Hayward,
  Phys.\ Rev.\  D {\bf 53}, 1938 (1996)
  [arXiv:gr-qc/9408002].


\bibitem{hideki}
  H.~Maeda and M.~Nozawa,
  Phys.\ Rev.\  D {\bf 77}, 064031 (2008)
  [arXiv:0709.1199 [hep-th]].


\bibitem{nozawa}
  M.~Nozawa and H.~Maeda,
  Class.\ Quant.\ Grav.\  {\bf 25}, 055009 (2008)
  [arXiv:0710.2709 [gr-qc]].




\bibitem{NM}
  M.~Nozawa and H.~Maeda,
  Class.\ Quant.\ Grav.\  {\bf 23}, 1779 (2006)
  [arXiv:gr-qc/0510070].

\bibitem{Hayward1993}
  S.~A.~Hayward,
  Phys.\ Rev.\  D {\bf 49}, 6467 (1994).


\bibitem{HIW}
  S.~Hollands, A.~Ishibashi and R.~M.~Wald,
  Commun.\ Math.\ Phys.\  {\bf 271}, 699 (2007)
  [arXiv:gr-qc/0605106].


\bibitem{Chan:1995fr}
  K.~C.~K.~Chan, J.~H.~Horne and R.~B.~Mann,
  Nucl.\ Phys.\  B {\bf 447}, 441 (1995)
  [arXiv:gr-qc/9502042].

\bibitem{Yazadjiev:2005du}
  S.~S.~Yazadjiev,
  Class.\ Quant.\ Grav.\  {\bf 22}, 3875 (2005)
  [arXiv:gr-qc/0502024].

\bibitem{soda}
  C.~Charmousis, B.~Gouteraux and J.~Soda,
  Phys.\ Rev.\  D {\bf 80}, 024028 (2009)
  [arXiv:0905.3337 [gr-qc]].


\bibitem{TM}
  T.~Torii and H.~Maeda,
  Phys.\ Rev.\  D {\bf 71}, 124002 (2005)
  [arXiv:hep-th/0504127];
  Phys.\ Rev.\  D {\bf 72}, 064007 (2005)
  [arXiv:hep-th/0504141].


\bibitem{Brill}
  D.~R.~Brill, G.~T.~Horowitz, D.~Kastor and J.~H.~Traschen,
  Phys.\ Rev.\  D {\bf 49}, 840 (1994)
  [arXiv:gr-qc/9307014].


\end{thebibliography}
\end{document}